\documentstyle[12pt,aaspp4,epsfig]{article}

\begin{document}

\received{12 April 1999}
\accepted{26 July 1999}
\journalid{125}{December 1999}

\slugcomment{To appear in ApJS.}

\lefthead{Preece et al.}
\righthead{BATSE Spectroscopy Catalog}

\title{The BATSE Gamma-Ray Burst Spectral Catalog. I. \\
High Time Resolution Spectroscopy of Bright Bursts \\
using High Energy Resolution Data}

\author{R.~D. Preece, M.~S. Briggs, R.~S. Mallozzi, G.~N. Pendleton, and 
W.~S. Paciesas}
\affil{Dept. of Physics, University of Alabama in Huntsville,
    Huntsville, AL 35899}

\and

\author{D.~L. Band}
\affil{Center for Astrophysics and Space Sciences, Code 0424, 
    University of California at San Diego, La Jolla, CA 92093}

\begin{abstract}
This is the first in a series of gamma-ray burst spectroscopy catalogs from 
the Burst And Transient Source Experiment (BATSE) on the {\it Compton Gamma 
Ray Observatory}, each covering a different aspect of burst phenomenology. 
In this paper, we present time-sequences of spectral fit parameters for 156 
bursts selected for either their high peak flux or fluence. All bursts have at 
least eight spectra in excess of $45 \sigma$ above background and span burst 
durations from 1.66 to 278 s. Individual spectral accumulations are typically 
128 ms long at the peak of the brightest events, but can be as short 
as 16 ms, depending on the type of data selected. We have used mostly 
high energy resolution data from the Large Area Detectors, covering an energy 
range of typically 28 -- 1800 keV. The spectral model chosen is from a small 
empirically-determined set of functions, such as the well-known `GRB' function, 
that best fits the time-averaged burst spectra. Thus, there are generally three 
spectral shape parameters available for each of the 5500 total spectra: a 
low-energy power-law 
index, a characteristic break energy and possibly a high-energy power-law index. 
We present the distributions of the observed sets of these parameters and comment 
on their implications. The complete set of data that accompanies this paper is 
necessarily large, and thus is archived electronically at: 
\verb+http://www.journals.uchicago.edu/ApJ/journal/+.
\end{abstract}


\section{Introduction}

Gamma-ray bursts (GRBs) have recently attracted considerable attention in 
the literature with the realization that burst sources are most likely 
associated with galaxies at cosmological distances. With the observation of GRB 
afterglows at many different wavelengths, there is an emerging picture of 
what happens {\em after} a burst is over. It is still unclear 
what, exactly, the burst is. For this reason, comprehensive studies of 
burst properties are necessary. We present here a  
spectroscopy catalog, complete as of Sept. 23, 1998, describing the 
high time-resolution spectral behavior of bright bursts as observed with the 
Burst And Transient Source Experiment (BATSE) on board the {\it Compton Gamma 
Ray Observatory}. 

Some results based upon these data have been reported elsewhere 
(\cite{kargatis96,crider97,preece98a,preece98b}). The first use of these data 
was to determine the low-energy behavior in peak outbursts in some GRBs as a 
function not of time, but of the integrated photon fluence (\cite{kargatis96}). This 
work was subsequently expanded upon (\cite{crider97}), in support of a saturated 
synchrotron self-Compton emission model for bursts. A detailed study of the temporal 
behavior of the high-energy power-law index made heavy use of spectral fit data for 
a large subset of the bursts from the catalog presented here (\cite{preece98a}). 
It was discovered that although bursts evolve from hard to soft in the high-energy 
power law as they do in the other spectral parameters (\cite{ford95}), this 
evolution is not directly correlated with that of any other parameter. Finally, 
the entire set of spectral fits has been used to rule out the synchrotron shock 
model as the emission mechanism for $\sim 30\%$ of all bursts, by a direct 
comparison between the observed distribution of the low-energy power-law indices 
with the predicted limiting value of $-2/3$ (\cite{preece98b}).

In this {\it Paper}, we bring together references to many aspects of the BATSE 
burst spectral observations that have heretofore been scattered throughout the 
literature. In the next section, we describe the two instruments that make up 
BATSE: the Large Area Detectors (LADs) and the Spectroscopy Detectors (SDs). In 
addition, the energy calibration procedures and their limitations are discussed.
In \S 3, the methodology employed to produce this catalog is explained. This includes 
event and data type selection as well as the spectral models used. The format 
of the catalog dataset is covered in detail in \S 4. Finally, in \S 5 we discuss 
some of the results that can be obtained using these data, by presenting the 
distributions of the fitted spectral shape 
parameters for the entire set of spectra. These can be a valuable tool 
for determining the validity of proposed physical emission models. As this 
catalog represents a very high level of data reduction, it is our hope that the 
data files that accompany this paper will be a valuable resource to the 
entire burst community.

\section{The BATSE Instrument}

\subsection{Overview}

BATSE was conceived of primarily as an all-sky gamma-ray transient monitor. 
For this reason, it consists of eight separate NaI-based detector modules located 
on all corners of the {\it Compton Observatory} spacecraft; four are on the `top', 
as defined by the placement and fields of view of the pointed 
gamma-ray instruments COMPTEL and EGRET, and four are on the bottom. To achieve the 
purpose of gathering information for the largest possible set of bursts, the LADs 
are constructed as a large, thin, flat collection area (2000 cm$^2$ $\times$ 
1.27 cm) with roughly a cosine angular response. To ensure maximal sky coverage 
for all eight detectors, the flat faces of the detectors are oriented parallel 
to the faces of a octahedron. 
The independent detectors' response is what allows for localization of 
bursting sources on the sky with an accuracy of $\sim 2^\circ$ (\cite{briggs99}).

The SDs were added to BATSE when a separate instrument for this purpose was not 
selected for inclusion on the {\it Compton Observatory}. They are smaller and 
thicker (127 cm$^2$ $\times$ 7.2 cm) than the LADs to maximize photon capture 
and energy resolution and thus have a more uniform angular response. Only 
at very low energies (5 to 15 keV) is the angular response more like a cosine, due 
to a circular beryllium window on the face of each SD. Each of the 8 BATSE modules 
consists of one LAD and one SD. The principal axis of each SD is offset in angle 
from the normal vector to the face of the associated LAD by $18^{\circ}.5$. The 
direction of the SD offset angle relative to the associated LAD is toward the 
midplane of the spacecraft, away from the axis defined by the two {\it Compton} 
pointed instruments.

\subsection{Energy Coverage}

As a result of their different purposes, the two types of detectors have different 
energy capabilities. The LADs are most useful if the energy range they observe is 
constant over time. This conforms to their design for long-term monitoring of 
transient gamma-ray sources. Thus, they have been gain-stabilized throughout the 
mission. The nominal energy coverage is 25 to 1800 keV, with some small 
variations between the detectors. For the purpose of triggering on burst-like 
events, the pulse heights of detected counts are divided into four broad channels 
by on-board fast discriminators. These are roughly 25 -- 50, 50 -- 100, 100 -- 
300, and $> 300$ keV. The on-board trigger can consist of any combination of these 
four channels in three timescales: 64, 256 and 1024 ms, but for the majority of 
the mission, the channel $2+3$, or 50 -- 300 keV, triggers has been used, since this 
combination is empirically found to be optimal for GRBs. It is an open question 
whether the burst samples obtained under different trigger criteria have differing 
spectral properties (\cite{lloyd99,harris98}). 

For LAD spectroscopy studies, the shaped pulse corresponding to the detection of a 
specific photon energy loss in the instrument is accumulated into one of 128 
quasi-logarithmic energy-loss channels (created from 2752 linear channels) by the 
on-board pulse-height analyzer (PHA), according to the pulse intensity. Given the 
moderate energy resolution of the LADs (an average of 19.6\% at 511 keV), this 
division oversamples the energy coverage. An energy resolution element can be 
defined to be equal to the FWHM of the detector energy resolution at a given 
energy. The typical energy range of the LADs is spanned by twenty energy 
resolution elements. The PHA output can be further rebinned 
into 16 channels according to a lookup table that is configured from the ground. 
Typically, the lookup table is constructed to sample the PHA output at roughly the 
detector resolution, resulting in the medium energy resolution of the MER/CONT data 
types (see below). However, at various 
points in the mission there was a need to observe one of several soft transient 
events at a higher resolution in the low-energy channels, so the table was modified 
accordingly. This is important for burst studies in that the energy sampling may 
contribute to the uncertainty in fitted spectral parameters when such data are used. 
No matter what the data type, a few of the lowest channels are not useful 
for spectroscopy, as they sample pulse heights that fall below the 
lower-level discriminator (LLD), an electronic cut-off that eliminates 
noise-produced counts.

For the SDs, all data are binned into 256 quasi-linear PHA channels. The lowest 
four of these, which would ordinarily contain residual noise counts from below the 
LLD, are replaced by data that serve the same purpose as the four LAD discriminators 
but have a much broader energy coverage, starting from one-half the LLD energy. 
The SD discriminator data are also available on a 2048 ms timescale. 
Unlike the LADs that have constant energy coverage over time, the SDs are separately 
configurable by several ground-commanded flight software parameters that have 
changed over time. The LLD setting determines the 
lowest usable PHA channel and the photo-multiplier tube (PMT) voltage sets 
the gain, or mapping, between PHA channels and energies of counts recorded in each 
channel. The higher the PMT voltages, the lower the energy coverage will 
be at any given LLD setting. The `nominal' ($1 \times$) gain setting places the 
511 keV annihilation feature in background spectra at linear PHA channel \#100 (the 
gain response is roughly linear). In this scale, a $4 \times$ gain setting nearly 
matches the energy coverage of the LADs. With low LLD and high gain ($8 \times$), 
the SD energy threshold is at $\sim 10$ KeV. 
The BATSE instrumental response functions are averaged over detector azimuth angle 
(Pendleton et al. 1995). The missing azimuthal dependence becomes significant 
when the zenith angle exceeds about $60^\circ$, so we restrict our analysis to 
cases with smaller angles.

\subsection{Energy Calibration and Instrumental Response}

The energy calibration of the two different BATSE detectors has been described 
elsewhere (\cite{preece98a,band92}). Residual non-linearities in the SD PHA output 
renders the first ten channels above the LLD unusable, although in many cases, there 
may be overlapping coverage from other detectors that may observe the same event. 
This omission stills allows for roughly two decades of 
coverage in each detector, which can be extended with the inclusion of the SD 
discriminator data below the LLD (\cite{preece96}), although we have not done this 
in the present catalog. The relative calibration between energy-integrated spectra 
for the LADs and the SDs has an apparent offset of about 10\%, partly due to the 
modelling of an aluminum honeycomb support structure for the LAD charged-particle 
detectors as a solid sheet of aluminum with an equivalent 
mass-thickness in the LAD response function. Work is now in progress to address this 
issue; however, the relative calibration becomes most important in joint fits of data 
from each detector type, which also have not been done here. 
In addition, if there were a method for determining the distance to each source from 
the flux, the absolute calibration (determining the difference between the observed 
flux and a standard reference, such as the Crab, for example) would be crucial. 
Currently, only a few bursts have had their red-shifts measured, although this will 
likely improve in the future.

\begin{figure*}[t!]
\centerline{\epsfig{file=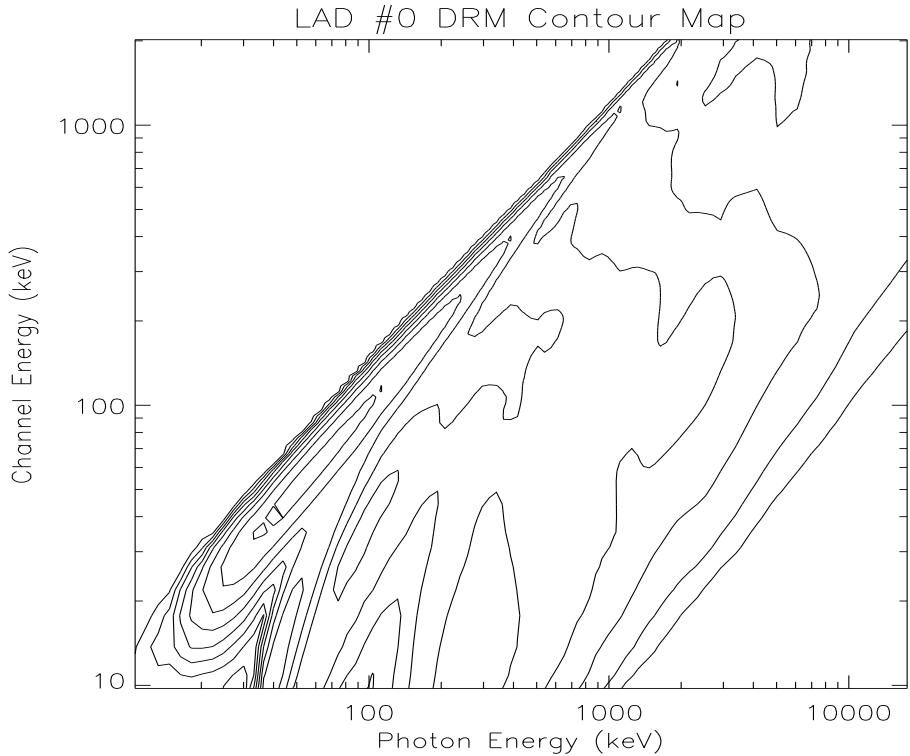,width=12cm,height=10cm}}
\caption{A typical BATSE LAD response matrix with equal levels 
of effective area indicated by 8 logarithmically-spaced contour lines, between 
0 and the maximum of 63 cm$^2$ keV$^{-1}$. The response to a photon of any given 
energy incident upon the detector can be traced by a vertical slice through the 
contour map at that energy. The photopeak is visible along the diagonal, while 
the iodine K-escape feature shows up along the bottom at 33 keV. 
\label{drm_fig}}
\end{figure*}

\begin{figure*}[t!]
\centerline{\epsfig{file=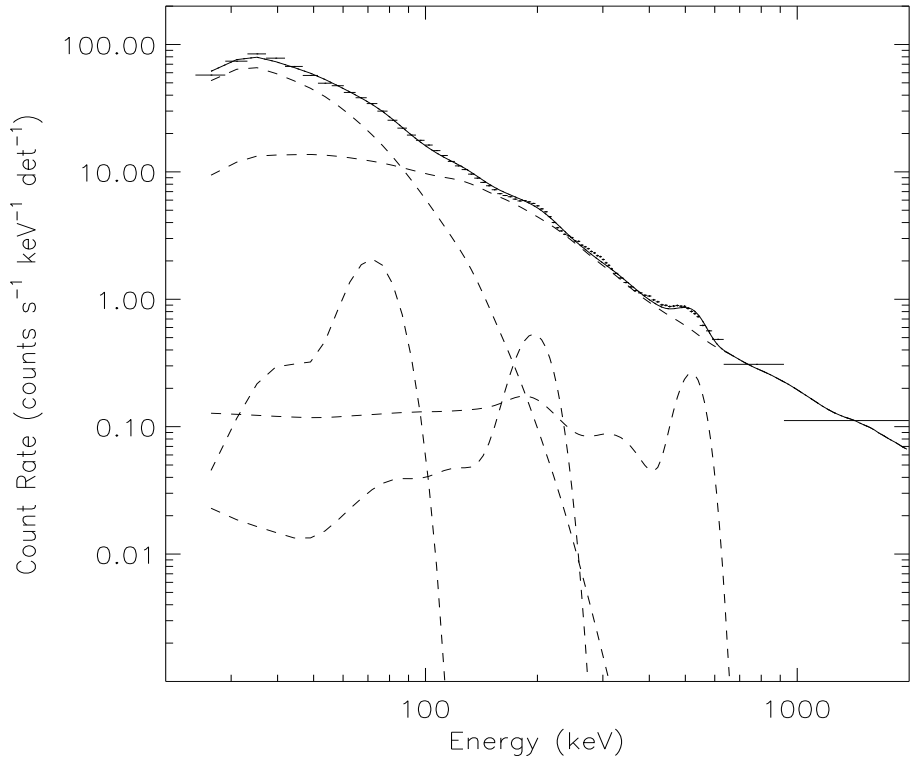,width=12cm,height=10cm}}
\caption{Fitted typical HER background spectrum for TJD 10459 
from 34536.2 to 34831.1 s UT, showing five components ({\it dashed}). The 
model includes a low-energy `thermal' portion, plus a $\sim -1$ spectral index 
high-energy power law and background lines at 511 keV, $\sim 200$ keV and 68 keV. 
The data are fit using these photon models propagated through the detector response 
matrix shown in Fig. \ref{drm_fig}. Note that some of the background continuum 
counts are due to particles, not gamma-ray photons.
\label{high_quality_spectrum}}
\end{figure*}

The description of the BATSE detector response matrices was presented by 
\cite{pendleton95}, while \cite{briggs96} reviewed the role of the detector 
characteristics in the forward-folding technique. Of interest in the current 
work is the nature of the 
off-diagonal contributions. The direct response (`photopeak') comes from 
photoelectric absorption of photons that interact in the body of the detector. 
Beside the photopeak, there is a large contribution generated by the iodine 
K-shell photon cross-section edge at $\sim 33$ keV. The lower cross-section 
below the K-shell ionization energy allows some photons to escape the detector. 
If the escaping photon is the X-ray florescence product of the photoelectric 
absorption of a higher-energy photon, a lower total energy will be recorded for 
the original photon. This effect can be seen in Figure \ref{drm_fig} as a 
broad second peak in the response that diverges from the photopeak at low 
energies. 
As this feature is imposed on the response by physics, it can serve 
as a convenient marker in energy, to aid in calibration of the gain. 
Photons that scatter into the detector from 
various sources, including the detector housing, the spacecraft, and the Earth's 
atmosphere, as well as those that Compton scatter out of the detector, make a 
broad lower-diagonal contribution from the higher energy external photons. These 
effects can be best seen in the counts predicted for photons of a single energy
interacting with the detector, as in Figure \ref{high_quality_spectrum} for 
the background spectral line features at 68, $\sim$ 200, and 511 keV. 
The various components of the response have differing energy 
dependencies that change with the viewing angles from the detector to the source 
and the Earth. Deconvolution of the count spectra becomes an exercise in the 
redistribution of the observed counts according to an assumed photon model; it 
is not unique.

The spacecraft flies in a roughly circular, inclined, and low-Earth orbit that 
creates a time- and energy-dependent 
background environment. In most cases, this background varies smoothly with 
time, so the separation of source and background counts is straightforward. 
However, there are a number of low-energy transient sources that contribute a 
considerable amount of noise when they are active, such as Sco X-1, the Vela pulsar 
and especially Cygnus X-1, that can emit weak, soft, flickering pulses at times of 
intense outburst. Such sources are most difficult for the SD data spectral analyses, 
as they are typically softer than the LAD energy bandpass, so they are not a concern 
here, except for those cases where SD data alone were available. 

\section{Methodology}

\subsection{Burst Selection Method}

In order to obtain a reasonable time-resolved picture of the spectral evolution in 
individual bursts, we consider bright bursts only in this catalog. We chose bursts 
to analyze using peak flux and total fluence limits, where these were determined 
by the methods described in the BATSE burst catalogs (\cite{cat1b,cat3b,cat4b}). By 
selecting bursts that satisfied either one of these criteria, we take maximum 
advantage of the time-to-spill spectral 
accumulation for the PHA data. For example, bursts with large peak flux may not have 
a large enough total fluence to be included in the sample by that criterion alone, 
but may still have a considerable number of spectra with sufficient counts to be 
useful for spectroscopic analysis. The total fluence selection criterion is 
$4 \times 10^{-5}$ erg cm$^{-2}$ integrated over all energies ($> 20$ keV). For 
peak flux, the catalog value must exceed 10 ph cm$^{-2}$ s$^{-1}$ on the 1024 ms 
integration timescale in the BATSE trigger energy band (50 -- 300 keV). Bursts 
were selected that had at least one of these catalog values available, to avoid any 
bias. 

The total duration of burst emission is selected by human judgement, usually 
starting with the first spectrum after the trigger, since for bright events, only 
unusual circumstances will allow any significant emission to be undetected before 
the trigger. Once the selection has been made, the spectra are rebinned in 
time according to a signal-to-noise ratio (S/N) criterion, in order to have a 
sufficient count rate in each spectrum so that the spectral model parameters could 
be determined with reasonable accuracy. To determine the S/N, the integrated 
background-subtracted count spectrum is compared with the total uncertainty, so the 
resulting measure is in units of standard uncertainty ($\sigma$). For LAD spectra 
over their total usable energy range, 
our minimum acceptable S/N was chosen to be 45, or roughly $2 \sigma$ per energy 
resolution element (discussed above) in a hypothetical flat LAD count spectrum. 
For most bright bursts, the peak spectra have count rates that are well in excess 
of $45 \sigma$, so it is typically the quiescent periods between peaks where 
spectra are binned together in this manner. SD data were used in some unusual cases 
where the LAD data were missing or unacceptable due to the presence of electronic 
distortions from overly high count rates. For these data, the S/N used to rebin 
the spectra in time was 15. Any bin at the end of the burst selection interval is 
likely to have less than $45 \sigma$, so it is dropped. We rejected bursts that had 
less than eight spectra total after the rebinning to ensure that enough spectra 
were available to track spectral evolution through each burst. After all the 
criteria have been applied to the entire BATSE data set from the beginning of 
the mission until Sept. 23, 1998, 156 bursts are available for this catalog.

\begin{figure*}[t!]
\centerline{\epsfig{file=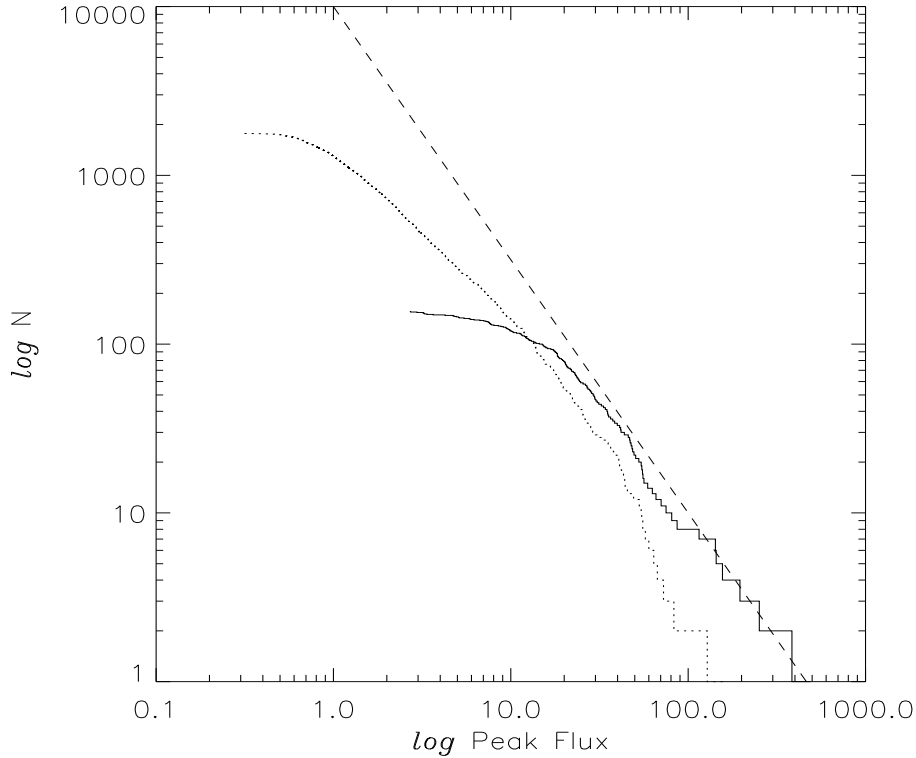,width=12cm,height=10cm}}
\caption{Peak flux distribution for the catalog bursts, compared 
with a $-3/2$ power law ({\it dashed}) as well as the 1772 BATSE 64 ms peak flux 
catalog values for the same period of time ({\it dotted} - the `Current' catalog: 
\protect \verb+http://www.batse.msfc.nasa.gov/batse/+). It is important to note that 
this 
set of bright bursts has an obvious selection bias, since they were selected on the 
basis of their total fluence or peak flux. In addition, there are peak fluxes based 
upon a 16 ms timescale included in our sample, so some values are brighter than 
they would be in the BATSE 64 ms peak flux data set.
\label{lognlogp}}
\end{figure*}

General information about the selected bursts is presented in Table 
\ref{cat_table}. Each burst is identified by the BATSE catalog name and trigger 
number (columns 1 \& 2), and several global properties are given that characterize 
each event, such as the number of spectra fit (column 6), the time interval chosen 
for fitting (7), and the peak flux and total fluence (8 \& 9) from the spectral 
fits. These last two values may differ from those in the official BATSE burst 
catalogs; the most important difference in the peak flux values is that there is no 
common time interval over which all of them are determined, since they depend 
upon the time-to-spill intervals in the spectral data. As such, they are 
given here as a rough guide of relative intensity. Indeed, the peak flux 
distribution (Figure \ref{lognlogp}) suffers from the bias of selecting bursts 
based upon total fluence or peak flux, rather than peak flux alone, as can be seen 
by the deficit in low peak flux bursts below 20 photons s$^{-1}$ cm$^{-2}$. 
The fluences have been 
calculated by integrating the individual spectral model fits over the total 
fitted energy range (typically 25 -- 1800 keV for LAD data) and summing 
over all spectra (as binned) in the time interval selected, which 
generally differs from the interval chosen for the BATSE burst catalog fluences. 
By using the best-fit model for each spectrum, this value represents a 
fair estimate of the fluence, at least for the $> 45\sigma$ portion of the 
burst history. The uncertainties associated with these two values were omitted 
for clarity and may be found in the online version of the catalog. The other 
columns of Table \ref{cat_table} refer to specifics of the data and spectral 
model selection, and will be described below.

\begin{deluxetable}{ccccccrrcc}
\tabcolsep=4pt
\scriptsize
\tablecaption{General Characteristics of Catalog Bursts. \label{cat_table}}
\tablewidth{0pt}
\tablehead{
  \colhead{Burst}
& \colhead{Trigger}  
& \colhead{Data}   
& \colhead{Detector} 
& \colhead{Model}  
& \colhead{No. of} 
& \multicolumn{2}{c}{Time Interval (s)} 
& \colhead{Peak Flux\tablenotemark{d}}     
& \colhead{Fluence\tablenotemark{d}}  
\nl
  \colhead{Name\tablenotemark{a}}   
& \colhead{Number}   
& \colhead{Type} 
& \colhead{Number\tablenotemark{b}} 
& \colhead{Used\tablenotemark{c}}  
& \colhead{Spectra} 
& \colhead{Start}  
& \colhead{Stop}
& \colhead{(ph~cm$^{-2}$~s$^{-1}$)}     %
& \colhead{(erg~cm$^{-2}$)}    
\nl %
  (1) 
& (2) 
& (3) 
& (4) 
& (5) 
& (6) 
& \multicolumn{2}{c}{(7)}  
& \multicolumn{1}{c}{(8)} 
& (9)
}
\startdata
3B 910421&     105&HERB&7&GRB& 14&     0.064& 10.304&    27.6& 8.2E$-$6\nl
3B 910425&     109&CONT&4&GRB& 14&   $-$16.384& 53.248&     4.7& 4.8E$-$5\nl
3B 910503&     143&HERB&6&GRB& 27&     0.704&  4.800&   142.7& 7.9E$-$5\nl
3B 910522&     219&MER&456&BPL& 51&   105.499&135.963&    33.1& 3.5E$-$5\nl
3B 910601&     249&HERB&2&GRB& 15&     0.0& 17.856&    28.1& 2.1E$-$5\nl
3B 910619&     394&HERB&1&GRB& 34&     0.0& 44.480&     7.6& 4.1E$-$5\nl
3B 910627&     451&HERB&4&GRB& 16&     0.0& 11.968&    21.9& 1.5E$-$5\nl
3B 910717&     543&HERB&4&GRB& 10&     0.0&  6.144&    16.4& 8.3E$-$6\nl
3B 910807&     647&HERB&0&BPL& 23&     0.0& 28.288&    10.2& 2.6E$-$5\nl
3B 910814C&     676&HERB&2&GRB& 18&     0.0& 54.592&     7.8& 3.0E$-$5\nl
3B 910814&     678&HERB&2&BPL& 37&     0.0& 29.376&    22.9& 7.8E$-$5\nl
3B 911031\tablenotemark{\dagger}&     973&HERB&3&GRB& 37&     0.0& 33.728&    12.1& 3.0E$-$5\nl
3B 911118\tablenotemark{\dagger}&    1085&HERB&4&BPL& 50&     0.0& 13.696&    49.0& 5.6E$-$5\nl
3B 911126&    1121&HERB&4&GRB& 28&    18.688& 29.888&    19.7& 2.0E$-$5\nl
3B 911127&    1122&HERB&1&GRB& 36&     0.0& 28.672&    18.3& 2.4E$-$5\nl
3B 911202&    1141&HERB&7&GRB& 31&     0.064& 17.856&    22.1& 4.2E$-$5\nl
3B 911209&    1157&HERB&1&GRB& 18&     0.128& 23.360&    19.1& 1.8E$-$5\nl
3B 920210&    1385&HERB&5&BPL& 35&     0.0& 48.768&     7.5& 4.1E$-$5\nl
3B 920226\tablenotemark{\dagger}&    1440&HERB&3&GRB& 16&     9.728& 17.920&    20.4& 1.3E$-$5\nl
3B 920311&    1473&HERB&5&GRB& 51&     3.392& 23.168&    46.5& 8.3E$-$5\nl
3B 920315&    1484&HERB&3&BPL&  8&     0.0& 20.096&    20.7& 5.7E$-$6\nl
3B 920406&    1541&SHERB&2&GRB& 20&    61.120& 87.232&    41.5& 6.3E$-$5\nl
3B 920513&    1606&CONT&3&GRB& 28&    12.288&102.400&     9.3& 5.8E$-$5\nl
3B 920525&    1625&HERB&4&GRB& 35&     4.160& 19.712&    54.6& 6.4E$-$5\nl
3B 920622\tablenotemark{\dagger}&    1663&HERB&4&GRB& 60&     0.0& 24.384&    39.0& 1.0E$-$4\nl
3B 920627&    1676&SHERB&2&GRB& 29&     0.064& 38.080&    12.4& 2.4E$-$5\nl
3B 920711&    1695&HERB&7&SBPL& 28&     0.0& 36.288&    35.9& 1.1E$-$4\nl
3B 920718&    1709&HERB&7&GRB& 13&     0.128&  5.056&    29.3& 9.9E$-$6\nl
3B 920723&    1721&HERB&3&BPL& 32&     0.064& 30.464&    14.0& 2.6E$-$5\nl
3B 920902&    1886&HERB&5&GRB& 30&     0.064& 14.592&    29.5& 4.5E$-$5\nl
3B 921003&    1974&HERB&2&GRB& 21&     0.0&  9.536&    13.2& 1.3E$-$5\nl
3B 921009&    1983&HERB&2&GRB& 54&     0.064& 29.184&    34.6& 6.3E$-$5\nl
3B 921015&    1989&CONT&4&BPL& 31&   110.592&350.208&     6.0& 4.4E$-$5\nl
3B 921022&    1997&HERB&2&GRB& 22&     0.0& 45.824&     9.9& 2.0E$-$5\nl
3B 921118&    2061&HERB&4&BPL& 16&     0.0& 50.560&     3.2& 2.2E$-$5\nl
3B 921123&    2067&HERB&1&GRB& 44&    12.160& 31.744&    29.8& 5.7E$-$5\nl
3B 921207\tablenotemark{\dagger}&    2083&HERB&0&GRB& 36&     0.0& 14.336&    88.3& 4.9E$-$5\nl
3B 921209&    2090&HERB&1&GRB& 17&     0.0& 13.888&    15.8& 1.2E$-$5\nl
3B 921230&    2110&MER&57&COMP& 29&    $-$2.048& 31.774&     7.5& 2.7E$-$5\nl
3B 930106&    2122&HERB&6&GRB& 12&     0.064& 73.856&     2.7& 1.9E$-$5\nl
3B 930120&    2138&MER&012&SBPL&125&    66.715&111.515&    17.3& 3.9E$-$5\nl
3B 930201&    2156&MER&137&SBPL&250&    $-$3.072&173.056&    37.2& 1.5E$-$4\nl
3B 930405&    2286&HERB&6&GRB& 26&     0.0& 25.152&    15.4& 2.4E$-$5\nl
3B 930425&    2316&HERB&1&GRB& 33&     0.0& 29.440&     7.6& 2.4E$-$5\nl
3B 930506&    2329&HERB&3&GRB& 26&     0.0& 13.888&    73.5& 1.2E$-$4\nl
3B 930916&    2533&HERB&3&GRB& 43&     0.0& 39.488&    13.7& 7.1E$-$5\nl
3B 930922&    2537&HERB&1&GRB& 23&     0.064&  4.864&    67.3& 1.6E$-$5\nl
3B 931008&    2571&MER&236&BPL& 43&     0.030&163.870&     7.8& 5.1E$-$5\nl
3B 931026&    2606&CONT&7&GRB& 18&    11.264& 97.280&     3.8& 3.1E$-$5\nl
3B 931103&    2617&HERB&5&GRB& 23&     0.064& 18.688&    15.2& 2.5E$-$5\nl
3B 931126&    2661&HERB&1&GRB& 18&     0.0& 12.928&    26.9& 2.9E$-$5\nl
3B 931204&    2676&HERB&1&GRB& 59&     0.0& 15.808&    50.0& 8.1E$-$5\nl
3B 940206&    2798&MER&13&SBPL& 65&     0.030& 68.126&    48.2& 1.4E$-$4\nl
3B 940210&    2812&HERB&6&BPL& 24&     0.0& 29.056&    11.4& 2.0E$-$5\nl
3B 940217&    2831&HERB&0&GRB& 32&     7.296& 33.664&    20.3& 7.4E$-$5\nl
3B 940228&    2852&HERB&7&BPL& 25&     0.0& 38.656&     9.4& 3.6E$-$5\nl
3B 940301&    2855&HERB&4&GRB& 37&     0.0& 43.264&    12.6& 6.5E$-$5\nl
3B 940302&    2856&CONT&1&GRB& 42&    $-$3.072&152.576&    23.0& 2.2E$-$4\nl
3B 940319&    2889&HERB&4&GRB& 10&     0.064& 63.488&     5.8& 3.6E$-$5\nl
3B 940323&    2891&MER&45&BPL& 36&    $-$2.048& 24.350&    18.6& 4.3E$-$5\nl
3B 940414&    2929&HERB&4&GRB& 36&     0.0& 46.208&     9.9& 4.6E$-$5\nl
3B 940429&    2953&HERB&3&GRB& 33&     0.0& 25.088&    26.7& 2.7E$-$5\nl
3B 940526B&    2993&HERB&3&COMP&  9&     0.0& 28.800&     4.8& 2.1E$-$5\nl
3B 940526&    2994&HERB&1&BPL& 44&     3.008& 27.264&    26.0& 5.0E$-$5\nl
3B 940529&    3003&HERB&0&GRB& 13&     0.0& 35.584&     5.1& 1.9E$-$5\nl
3B 940619&    3035&HERB&6&GRB& 13&     0.0& 62.016&     3.2& 1.9E$-$5\nl
3B 940623&    3042&HERB&1&GRB& 19&     0.128& 16.192&    12.1& 1.5E$-$5\nl
3B 940703&    3057&SHERB&5&BPL& 22&    27.456& 92.800&    47.0& 2.3E$-$4\nl
3B 940708&    3067&HERB&6&GRB& 25&     0.0&  7.744&    35.2& 3.0E$-$5\nl
3B 940810&    3115&HERB&3&GRB& 21&    10.944& 30.656&    18.3& 1.8E$-$5\nl
3B 940817&    3128&HERB&5&GRB& 43&    20.416& 48.000&    20.8& 6.1E$-$5\nl
3B 940826&    3138&HERB&6&GRB& 18&     7.808& 17.408&    25.1& 1.1E$-$5\nl
4B 940921&    3178&HERB&2&GRB& 22&     0.033& 24.768&    35.0& 5.4E$-$5\nl
4B 941008&    3227&HERB&5&GRB& 14&    82.176& 93.056&     8.7& 1.2E$-$5\nl
4B 941014&    3241&HERB&6&GRB& 37&    18.560& 44.736&    18.3& 3.0E$-$5\nl
4B 941017&    3245&MER&04&SBPL&182&     4.894& 86.558&    23.4& 1.6E$-$4\nl
4B 941020\tablenotemark{\dagger}&    3253&SHERB&5&GRB& 28&    15.680& 70.336&    20.5& 6.7E$-$5\nl
4B 941023&    3255&HERB&4&BPL& 14&     0.127& 32.320&    20.0& 2.0E$-$5\nl
4B 941121&    3290&HERB&4&GRB& 12&    34.304& 52.032&    18.6& 1.6E$-$5\nl
4B 941228&    3330&HERB&3&GRB& 14&     0.108& 59.584&     9.8& 2.2E$-$5\nl
4B 950104&    3345&HERB&1&GRB& 17&     0.033& 11.584&    11.5& 1.4E$-$5\nl
4B 950111&    3352&HERB&2&GRB& 28&     0.033& 42.560&     5.9& 2.6E$-$5\nl
4B 950208&    3408&HERB&6&GRB& 61&     0.090& 28.288&    24.0& 5.5E$-$5\nl
4B 950211&    3415&HERB&5&GRB& 25&     0.026& 54.592&    15.8& 2.9E$-$5\nl
4B 950301&    3448&CONT&3&BPL& 11&   180.224&337.920&     2.7& 2.3E$-$5\nl
4B 950305\tablenotemark{\dagger}&    3458&SHERB&4&GRB& 17&     0.033& 23.040&    11.1& 2.1E$-$5\nl
4B 950325&    3481&HERB&2&GRB& 21&    37.184& 47.616&    55.7& 2.7E$-$5\nl
4B 950401&    3489&HERB&5&GRB& 15&     0.116& 18.624&     9.6& 2.6E$-$5\nl
4B 950403&    3491&HERB&3&GRB& 52&     1.792& 17.344&    61.2& 4.4E$-$5\nl
4B 950403B&    3492&HERB&5&GRB& 26&     3.008&  8.960&   224.5& 4.3E$-$5\nl
4B 950425&    3523&HERB&6&BPL& 58&     0.033& 28.864&    47.2& 1.1E$-$4\nl
4B 950513&    3571&HERB&5&GRB& 22&     0.092& 36.352&    18.0& 2.3E$-$5\nl
4B 950522&    3593&HERB&2&BPL& 14&     0.026& 18.368&    11.5& 2.1E$-$5\nl
4B 950701&    3657&HERB&4&GRB& 26&     0.032&  9.600&    48.3& 2.3E$-$5\nl
4B 950701B&    3658&HERB&5&GRB& 30&     0.026& 11.392&    19.5& 2.0E$-$5\nl
4B 950804&    3734&HERB&4&GRB& 15&     0.147&  3.712&    55.0& 2.0E$-$5\nl
4B 950818&    3765&HERB&1&GRB& 31&    51.584& 73.280&    41.7& 3.4E$-$5\nl
4B 950909&    3788&CONT&3&GRB& 23&     0.0& 67.584&     8.3& 3.9E$-$5\nl
4B 951011&    3860&HERB&5&GRB& 16&     0.033& 29.760&     6.4& 2.9E$-$5\nl
4B 951016&    3870&HERB&5&GRB& 16&     0.026&  5.440&    30.5& 1.2E$-$5\nl
4B 951102&    3891&HERB&2&GRB& 21&    25.920& 42.368&    28.3& 1.3E$-$5\nl
4B 951203&    3930&HERB&0&GRB& 20&     0.032& 21.056&    17.6& 5.4E$-$5\nl
4B 951219&    4039&HERB&6&COMP& 19&     0.025& 43.008&    11.8& 3.7E$-$5\nl
4B 960114&    4368&SHERB&0&BPL& 25&    14.784& 27.200&   142.2& 1.2E$-$4\nl
4B 960124\tablenotemark{\dagger}&    4556&HERB&5&GRB& 23&     0.026&  4.992&    28.8& 1.6E$-$5\nl
4B 960201\tablenotemark{\dagger}&    4701&HERB&1&GRB& 21&     7.232& 26.304&    10.3& 2.1E$-$5\nl
4B 960321&    5299&SHERB&0&BPL& 46&    41.280& 69.440&    54.1& 4.5E$-$5\nl
4B 960322\tablenotemark{\dagger}&    5304&MER&57&GRB& 59&     0.024& 26.904&    23.5& 7.3E$-$5\nl
4B 960529&    5477&HERB&1&COMP& 20&     0.091& 10.048&    76.6& 3.5E$-$5\nl
4B 960605&    5486&MER&23&BPL& 46&    59.672& 89.368&    17.4& 5.9E$-$5\nl
4B 960607&    5489&HERB&1&GRB& 26&    29.888& 55.040&    14.9& 2.9E$-$5\nl
4B 960623&    5512&HERB&5&BPL& 13&     0.033& 29.952&     5.1& 1.8E$-$5\nl
4B 960807&    5567&HERB&0&GRB& 32&     0.090& 17.792&    45.4& 2.7E$-$5\nl
4B 960808&    5568&HERB&6&GRB&  8&     0.190&  5.632&    24.3& 1.9E$-$5\nl
5B 960831&    5591&MER&67&BPL& 49&    $-$9.216&166.912&     7.4& 5.6E$-$5\nl
5B 960924&    5614&SHERB&6&BPL& 27&     7.360& 13.120&   490.0& 1.5E$-$4\nl
5B 961001\tablenotemark{\dagger}&    5621&HERB&2&GRB& 25&     0.026& 10.304&    51.2& 2.9E$-$5\nl
5B 961009&    5629&HERB&6&BPL& 35&     0.033& 15.936&     9.5& 2.1E$-$5\nl
5B 961029\tablenotemark{\dagger}&    5649&MER&15&SBPL&351&     1.941&123.285&    55.6& 1.8E$-$4\nl
5B 961102&    5654&MER&15&GRB& 63&     0.030&100.126&     7.1& 6.0E$-$5\nl
5B 961202&    5704&HERB&0&GRB& 13&     0.030&  5.184&    64.4& 1.2E$-$5\nl
5B 970111&    5773&HERB&0&COMP& 57&     0.029& 21.824&    23.8& 4.4E$-$5\nl
5B 970201&    5989&MER&015&GRB& 41&     0.024&  1.688&   279.4& 1.0E$-$5\nl
5B 970202\tablenotemark{\dagger}&    5995&HERB&1&BPL& 60&     0.090& 21.632&    24.4& 7.3E$-$5\nl
5B 970223&    6100&HERB&6&BPL& 35&     0.090& 18.368&    35.3& 4.2E$-$5\nl
5B 970306&    6115&HERB&2&BPL& 15&     0.150&107.776&     3.3& 4.0E$-$5\nl
5B 970315&    6124&SHERB&2&BPL& 47&     0.026& 19.200&    40.6& 6.4E$-$5\nl
5B 970411&    6168&CONT&1&GRB& 17&   $-$12.288&102.400&    18.3& 8.2E$-$5\nl
5B 970420&    6198&HERB&4&GRB& 45&     0.045& 11.968&   145.0& 6.4E$-$5\nl
5B 970517&    6235&HERB&5&GRB&  9&     0.090&  7.232&    31.7& 1.5E$-$5\nl
5B 970616&    6274&HERB&1&BPL& 20&    42.240&127.040&    21.3& 3.9E$-$5\nl
5B 970807&    6329&HERB&4&GRB& 31&     0.045& 47.744&    12.6& 3.9E$-$5\nl
5B 970816&    6336&HERB&7&SBPL& 10&     1.984&  6.848&    29.9& 2.4E$-$5\nl
5B 970828&    6350&MER&57&COMP&113&     0.030& 96.030&    15.3& 7.4E$-$5\nl
5B 970831&    6353&HERB&0&GRB& 32&     0.033&123.008&     5.1& 5.0E$-$5\nl
5B 970919&    6389&MER&013&GRB& 28&   $-$14.016& 24.601&    12.2& 3.4E$-$5\nl
5B 970925&    6397&HERB&7&GRB& 19&     0.092& 35.584&    10.4& 2.4E$-$5\nl
5B 970930&    6404&HERB&6&GRB& 20&     0.058& 11.648&    42.0& 2.1E$-$5\nl
5B 971029&    6453&HERB&1&BPL& 45&     0.033&117.312&     7.0& 5.8E$-$5\nl
5B 971110&    6472&CONT&1&SBPL& 52&    $-$2.304&276.224&    23.0& 1.7E$-$4\nl
5B 971208&    6526&CONT&6&SBPL& 36&    $-$5.120&199.680&     3.5& 9.4E$-$5\nl
5B 980105&    6560&HERB&7&SBPL& 16&     0.030& 37.056&    13.7& 2.0E$-$5\nl
5B 980124&    6576&MER&046&SBPL& 55&     0.027& 45.595&    20.2& 5.4E$-$5\nl
5B 980125&    6581&HERB&0&SBPL& 17&    42.688& 64.768&    85.0& 2.9E$-$5\nl
5B 980203&    6587&MER&01&SBPL& 94&     0.030& 36.126&    57.0& 1.3E$-$4\nl
5B 980208\tablenotemark{\dagger}&    6593&HERB&3&SBPL& 32&     0.033& 29.632&    17.9& 3.7E$-$5\nl
5B 980225&    6615&HERB&4&SBPL& 16&    85.120&164.096&     3.2& 5.7E$-$5\nl
5B 980306&    6629&HERB&1&GRB& 12&   180.608&254.144&    12.9& 4.5E$-$5\nl
5B 980306B&    6630&HERB&3&SBPL& 17&     0.042& 20.992&    21.9& 2.2E$-$5\nl
5B 980329\tablenotemark{\dagger}&    6665&HERB&0&SBPL& 38&     0.026& 20.736&    26.9& 4.9E$-$5\nl
5B 980703&    6891&MER&13&SBPL& 27&   $-$34.816& 59.550&     4.8& 3.7E$-$5\nl
5B 980724&    6944&HERB&6&SBPL& 23&     0.090& 50.560&     8.9& 2.8E$-$5\nl
5B 980803&    6963&MER&26&SBPL& 31&    $-$2.048& 24.607&    30.9& 4.4E$-$5\nl
5B 980810&    6985&HERB&3&SBPL& 56&     0.033& 42.432&    33.6& 1.1E$-$4\nl
5B 980821&    7012&HERB&0&SBPL& 32&     0.091& 30.784&    38.0& 3.9E$-$5\nl
5B 980923&    7113&MER&37&SBPL&133&     0.024& 38.424&   167.3& 4.5E$-$4\nl
\tablenotetext{a}{Burst names are from \cite{cat3b} (3B Catalog) and \cite{cat4b} 
(4B), and continue with a `5B' prefix for bursts that occurred after the end of 
the 4B Catalog.}
\tablenotetext{b}{Entries with multiple detectors indicate that summed data
were used for the analysis.}
\tablenotetext{c}{Model names are described in the text.}
\tablenotetext{d}{Fluences are determined for the given spectral form and are 
integrated over the indicated time interval; peak fluxes have varying timescales. 
Thus, these differ from the 3B and 4B Catalog values.}
\tablenotetext{\dagger}{Overwriting event.}
\enddata
\end{deluxetable}
\clearpage

\subsection{Selection of Appropriate Data Type}

\subsubsection{Burst Data}

There are several data types available for use, each of which has its own advantages 
and disadvantages. Table \ref{datatypes} shows the differences among all the 
BATSE data types available for burst studies, in terms of energy and temporal 
resolutions. The time coverage for the high energy resolution PHA 
data types (HERB/SHERB) varies, depending on the burst count rate, 
but is never longer than the burst data accumulation period (Table \ref{burstaccum}). 
Faced with the array of data products, several criteria have to be set to ensure the 
maximum scientific return from the analyses. In view of the fact that the LAD 
collecting area is almost $16 \times$ that of the SDs, data from the LADs are 
preferred. For continuum studies, the higher count rate, combined with the moderate 
energy resolution, is sufficient to determine spectral model fit parameters with good 
precision. The HERB data type has the highest energy resolution and sub-second 
time resolution (depending on the burst intensity), so it is the primary data type 
used for this effort. If the HERB accumulation finishes before the end of the burst, 
or if the HERB time resolution is not sufficient to resolve significant 
features in the time history, MER may be substituted. If that is not available, 
then CONT data can be used, especially in cases where the burst is longer than the 
burst accumulation time (Table \ref{burstaccum}) or the MER duration. The 
discriminator (4-channel) data are not used for this catalog since they lack sufficient 
energy resolution. Finally, there are occasions where an appropriate LAD data type is 
unavailable, such as loss of data transmission due to telemetry gaps. If the event is 
intense enough that the difference in collecting area does not matter much, the 
SD PHA data type SHERB may be substituted. It is also preferred in very intense 
events where the available HERB memory fills up before the end of the outburst. For 
the brightest of these, electronic effects such as pulse pile-up precludes 
analysis. In these cases (only two during the 
period covered by this catalog), the SD data are the only recourse for accurate 
spectral deconvolution. Table \ref{cat_table} gives the chosen data type in column
3, and the corresponding detector number(s) in column 4. In particular, note from  
the `SHERB' entries in column 3 that SD data were used for only nine bursts 
out of the total sample.

\begin{deluxetable}{lcccccr}
\footnotesize
\tablecaption{BATSE Burst Data Types\label{datatypes}}
\tablehead{
\colhead{Data} & 
\colhead{Detector} & 
\colhead{\# Energy} & 
\colhead{\# Spectra} & 
\colhead{Resolution} &
\colhead{Detector} &
\colhead{Time} \nl
\colhead{Type} & 
\colhead{Type} & 
Channels & 
(\# Events)\tablenotemark{a} & 
(ms) & 
Subset & 
\colhead{Coverage}
}
\startdata
HERB & LAD & 128 & 128 & 128\tablenotemark{b} & DSELH\tablenotemark{c} & $\le$ DISCSC \nl
HER & LAD & 128 & --- & ($\sim 300$ s) & All & Background \nl
MER & LAD & 16 & 4096 & 16 \& 64\tablenotemark{d} & DSELB\tablenotemark{e} & 163.84 s \nl
CONT & LAD & 16 & --- & 2048 & All & Continuous\tablenotemark{f} \nl
DISCSC & LAD & 4 & Varies\tablenotemark{g} & 64 & DSELB\tablenotemark{e} & 
Fixed\tablenotemark{g} \nl
PREB & LAD & 4 & 32 & 64 & All & $-2.048$ s \nl
TTE & LAD & 4 & (32768) & 0.128 & DSELB\tablenotemark{h} & Varies \nl
DISCLA & LAD & 4 & --- & 1024 & All & Continuous\tablenotemark{f} \nl
SHERB & SD & 252 & 192 & 128\tablenotemark{b} & DSELH & $\le$ DISCSC \nl
SHER & SD & 252 & --- & ($\sim 300$ s) & All & Background \nl
STTE & SD & 256 & (65536) & 0.128 & DSELB\tablenotemark{h} & Varies \nl
DISCSP & SD & 4 & 192 & 2048, 128\tablenotemark{b} & All, DSELH & Continuous \nl
\tablenotetext{a}{(S)TTE: Individual time-tagged events in parentheses.}
\tablenotetext{b}{Minimum for time-to-spill, increases by 64 ms increments.}
\tablenotetext{c}{DSELH: 4 detectors with `highest' count rates, determined at trigger.}
\tablenotetext{d}{The change to 64 ms resolution is after the first 32.768 s.}
\tablenotetext{e}{DSELB: The 2--4 detectors with highest count rates, as 
determined at the time of the trigger (MER and DISCSC are summed over these 
detectors).}
\tablenotetext{f}{Can serve as background data.}
\tablenotetext{g}{Determined by the length of the data accumulation (see Table 
\protect\ref{burstaccum}).}
\tablenotetext{h}{Switches from the full set of detectors to the DSELB subset after 
approx.~30 ms.}
\enddata
\end{deluxetable}

\begin{deluxetable}{rrrr}
\footnotesize
\tablewidth{0pt}
\tablecaption{Burst Data Accumulation History\label{burstaccum}}
\tablehead{
\colhead{Date} & 
\colhead{Seconds} & 
\colhead{\# DISCSC} &
\colhead{Accumulation} \\
\colhead{(TJD)} & 
\colhead{of Day\tablenotemark{\dagger}} & 
\colhead{Spectra} & 
\colhead{Duration (s)}
}
\startdata
8362  &        0  & 3776  & 241.664 \nl
8366  &        0  & 2816  & 180.224 \nl
8367  &     5632  & 3776  & 241.664 \nl
8807  &    66200  & 2816  & 180.224 \nl
8810  &    78770  & 3776  & 241.664 \nl
8973  &    62418  & 8960  & 573.440 \nl
10091 &     6591  & 3776  & 241.664 \nl
10181 &    55298  & 8960  & 573.440 \nl
10464 &    71456  & 3776  & 241.664 \nl
10504 &    80943  & 8960  & 573.440 \nl
10983 &    71859  & 3776  & 241.664 \nl
\tablenotetext{\dagger}{Date and time at which the accumulation mode went into effect.}
\enddata
\end{deluxetable}

The high energy resolution PHA data types HERB and SHERB 
are accumulated in a time-to-spill mode, not uniformly sampled in time. As 
originally configured, the flight software would end a spectral accumulation 
after 64k counts (or any other preset value) had been recorded in one of the 
LAD detectors. The detector picked to have its count accumulations monitored 
for this purpose is the one that was determined to have the 
highest count rate at the burst trigger time. Spectral data are only accumulated 
for the detectors that had the four highest count rates at the trigger time. 
The second `brightest' detector records one spectrum for every two the brightest 
detector records, while the third and fourth brightest detectors record one spectrum 
for every two recorded in the second brightest. The minimum accumulation time 
is 128 ms; however, the incremental time resolution is 64 ms. A total of 128 
spectra can be accumulated for the LADs, while 192 are available for SD spectra. 
After half of the LAD spectra have been accumulated, the on-board CPU tests the  
elapsed time against a commanded time parameter. If the actual time is larger, 
a multiplication constant (usually 2), which can also be changed from the ground, 
is then applied to the spectral accumulation parameter, until the memory fills or 
a preset time runs out (this is the `burst accumulation time', which is also set 
from the ground -- see Table \ref{burstaccum}). The time resolution of the two BATSE 
detector types is identical until 
the LADs fill up the on-board memory, at which time the electronics again determine 
whether it should change the accumulation criteria for the remaining SD memory. 
In a later change of the flight software, the total counts minus a ground-commanded 
fraction of the background counts, rather than the total counts alone, is compared 
with the fixed total counts accumulation criterion for each spectrum. To obtain 
approximately the same peak accumulations as before this change, the accumulation 
parameter is also lowered from the previous value of 64k counts to accommodate 
for the reduction of the total count rate by the background fraction. In this 
manner, periods where the burst activity has effectively returned to background 
levels are accumulated into one spectrum, while active portions of the 
burst are sampled nearly as rapidly as previously. The intention is to provide 
fine time-resolution spectra throughout a long event, without having to change to 
longer accumulations to cover the end. 

\subsubsection{Continuous and Background Data}

Some cases exist where the selected burst in the catalog triggered the instrument 
during the data readout period (usually about 90 min) of a weaker event. In which 
case, data for the previous event not yet telemetered to the ground are overwritten 
in the BATSE memory by the new event; hence this class of events are called 
`overwrite' triggers. During the burst data readout, the 64 ms trigger threshold is 
set to either the larger of the highest recorded 64 ms count rate in the brightest 
detector during the burst data accumulation period or the largest of the previous 
event's three trigger thresholds. Given how the BATSE trigger operates, count rates 
in the {\em second} brightest detector must exceed the new 64 ms trigger threshold 
(the two other trigger timescales are disabled) before an overwrite event 
occurs. Thus, the overwriting event must be quite strong before it can trigger, and 
possibly some considerable portion of the emission may occur before the trigger. The 
PHA data do not have fine time resolution before the trigger, so an 
alternative data type (usually CONT) must be chosen. Overwriting bursts are 
indicated with a dagger ($\dagger$) in column 1 of Table \ref{cat_table}.

\begin{figure*}[t!]
\centerline{\epsfig{file=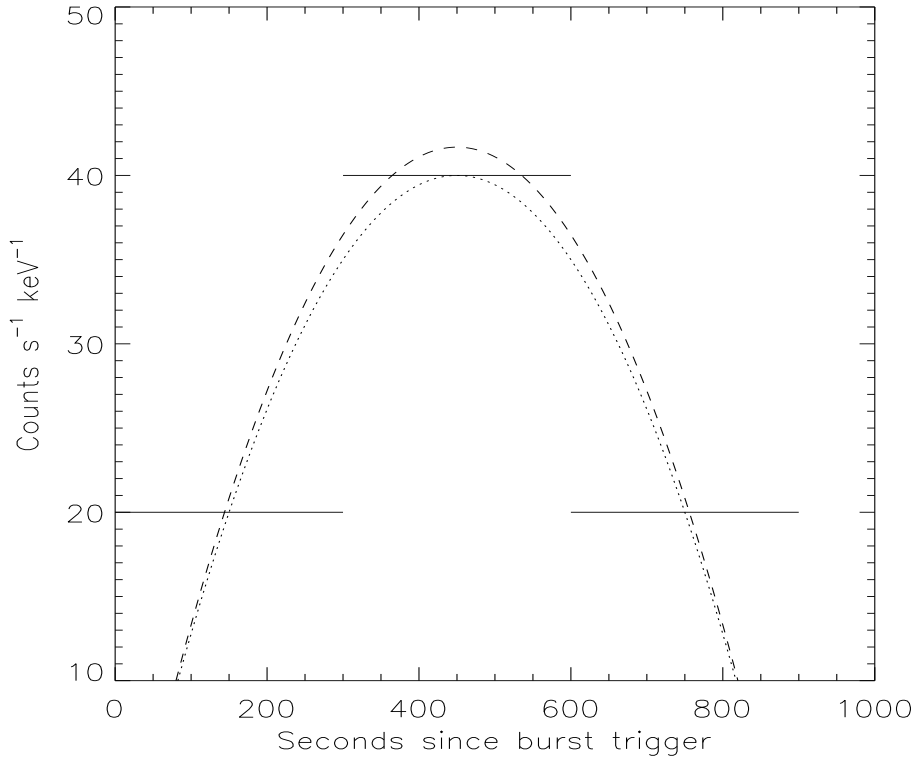,width=12cm,height=10cm}}
\caption{A simulated set of HER background histories, with a 
concave upwards behavior, showing two fits: one, a simple second-order polynomial 
in time using the average value for each rate fitted at the center, obviously 
underestimating the average rate ({\it dotted}), the other using integrated basis 
functions, where the average of the fitted model over the spectral accumulation 
time interval is the same as the data rate ({\it dashed}). In practice, a 
second-order fit to three background intervals would not be allowed.
\label{background}}
\end{figure*}

The background rates are available in corresponding data types for each `class' of 
energy resolution, as seen in Table \ref{datatypes}. For example, the 128 
channel HER data type is the appropriate background data for the HERB burst data 
type. Due to memory storage and telemetry limitations, the highest energy 
resolution background data are available in the lowest time resolution, typically 
$\approx 300$ s for HER and SHER. The result of this is that only long-term variations 
of the background rate can be modelled in most cases. For background modelling, 
low-order ($\le 4$) polynomial behavior is assumed. 
Because the length of the background accumulations of HER and SHER data approach
the timescale of background variations, the background model must predict the
average background rate of an interval, rather than the background rate at the
midpoint of the interval.
As an example, consider the case that the background rate during three
background accumulations is low, high, then low (Fig.~\ref{background}).
If we fit a second-order polynomial using points at the center of each
accumulation, then the estimated rate for any time during the center interval
will always be less than the true rate (Fig.~\ref{background}--{\it dashed line)}).
The observed rates are averages over the accumulation intervals,
so the background model should also be averaged over the intervals.
The correct fit is obtained by minimizing the differences between
the averages of the polynomial model and the observed rates
(Fig.~\ref{background}--{\it dotted line}). 
When they are available, we choose background spectra within 
$\pm 1000$ s relative to the trigger, so that at least three spectra before and 
after the trigger can be used to provide a background model. For cases 
where the burst emission is over before the burst data memory has been filled, 
the post-burst portion may be used as part of the background. It is better to 
estimate the background as close as possible to the beginning and the end of the 
burst; however, care must be 
taken so as not to include a faint, fading portion of the burst. If less than six 
total background spectra are available, the order of the background model must be 
decreased from the default of four, so that the available degrees of freedom does 
not fall to zero or below. The model is fit separately to each energy channel, and the 
resulting $\chi^2$ per degree of freedom examined for systematic problems. If any 
are found, the background selection is changed and re-fit until a suitable model 
is found.

\subsection{Spectral Models Used}

Due to the non-diagonal nature of the detector response function, the 
choice of spectral model to be used for fitting in some sense determines the 
outcome. Spectral deconvolution is never exact, since we are trying to guess 
an unknown function from a data set that has been passed through a filter that 
loses information. That is, we choose a candidate spectral form, with several 
parameters to be determined from the fit, and multiply it by the detector response 
function, which distributes photons at any given energy into counts data channels 
at all lower energies. The functional parameters are adjusted until a best fit 
is obtained. However, the best fit is only for the chosen model; there is no 
guarantee that a different model might not result in a better fit. Unfortunately, 
there is no best method to determine for each count spectrum the best possible 
functional form other than trying many different functions, optimizing each one 
over the available spectral parameters and choosing the one with the best $\chi^2$. 
In practice, we have drawn from a small set of `reasonable' spectral forms, all 
empirically determined through experience. We do make the 
assumption, however, that a single spectral model suffices to fit every spectrum 
in a single burst, an assumption that can be tested using the $\chi^2$ distribution. 
For each burst, the particular model used is given by a mnemonic (such as `GRB') in 
column 5 of Table \ref{cat_table} that is explained in the following discussion. 
The models and their parameters used in the preparation of this catalog are 
presented below as they appear in the data archive.

\subsubsection{The GRB Model}

The model typically chosen for spectral fitting of bursts is the empirical 
`GRB' function (\cite{band93}): 
\begin{eqnarray}
f_{\rm GRB}(E) & = & A (E/100)^{\alpha} \exp{(-E(2+\alpha)/E_{\rm peak})}\nonumber\\
{\rm if} \quad E & < & (\alpha-\beta)E_{\rm peak}/(2+\alpha) \equiv E_{\rm break} {\rm ,}\\
{\rm and} \quad f(E) & = & A \{(\alpha-\beta)
E_{\rm peak}/[100(2+\alpha)]\}^{(\alpha-\beta)} \nonumber\\
& & \exp{(\beta-\alpha)} (E/100)^{\beta}\nonumber\\
{\rm if} \quad E & \geq & (\alpha-\beta)E_{\rm peak}/(2+\alpha){\rm ,} \nonumber
\end{eqnarray}
where the four model parameters are 
\begin{enumerate}
\item  the amplitude $A$ in photons~s$^{-1}$~cm$^{-2}$~keV$^{-1}$,
\item  a low-energy spectral index $\alpha$, 
\item  a high-energy spectral index $\beta$,
\item  and a $\nu {\cal F}_{\nu}$ `peak' energy $E_{\rm peak}$.
\end{enumerate}
The parameter $E_{\rm peak}$ corresponds to the peak of the spectrum in $\nu 
{\cal F}_{\nu}$ only if $\beta$ is less than $-2$. Otherwise, the parameter value 
is identical to $E_{\rm break}$, which marks the lower boundary in energy of the 
high-energy power-law component characterized by $\beta$. The `true' $E_{\rm peak}$ 
when $\beta > -2$ lies at an unknown energy beyond the high end of the data, for 
867 out of the 5253 spectra that had a well-defined value of $\beta$ (16\%). 
The `pivot' energy of 100 keV serves as the energy at which the normalization of 
the function $A$ is determined. Several 
words of caution are in order here: the GRB model parameter $\alpha$ is an asymptote 
to the actual power-law slope realized by the data (\cite{preece98b}). The inherent 
curvature of the model may or may not not be well realized in the low-energy data, 
but there is no way the curvature can be adjusted arbitrarily by the spectral 
parameters. If all three parameters are well-determined, the curvature is fixed by 
the assumed functional form. There is another function that addresses this (see 
the SMPL model below), at the cost of an additional degree of freedom. 

\subsubsection{The COMP Model}

If the GRB model break energy lies above the highest energy available to the 
detector so that $\beta$ is ill-defined, or if $|\beta|$ is so large that is 
essentially infinite (and thus numerically unstable), we must substitute a 
related form of the model with the high-energy power-law omitted:
\begin{equation}
f_{\rm COMP}(E) = A (E/E_{\rm piv})^{\lambda} \exp{(-E(2+\lambda)/E_{\rm peak})}
\end{equation}
where the four model parameters are 
\begin{enumerate}
\item  the amplitude $A$,
\item  the $\nu {\cal F}_{\nu}$ `peak' energy $E_{\rm peak}$ in keV,
\item  a low-energy spectral index $\lambda$, 
\item  $E_{\rm piv} =$ a pivot energy in keV (always held fixed).
\end{enumerate}
For historical reasons, the resulting model is 
called `COMP', meaning Comptonized spectrum. However, the COMP functional form 
describes the physical thermal Compton spectrum only for the 
value $\alpha = -1$; whereas, in fitting, there is usually no such restriction. 
Again, the pivot energy determines the energy at which the normalization $A$ is 
calculated. For some soft spectra (not considered here), the value of the pivot 
greatly affects the numerical stability of the fit. If the peak energy is less 
than 10 KeV, for example, the normalization determined at 100 keV would be quite large 
and may cause a numerical overflow; in which case, the pivot should be changed 
to 10 keV. In the catalog, this and the corresponding parameters in the models 
below are always held constant throughout the series of fits in each burst 
at 100 keV.

\subsubsection{The BPL Model}

In cases where a sharp curvature of the spectrum results in an unacceptable 
value of $\chi^2$ for a fit to the intrinsically smooth GRB spectral form, a broken 
power-law (`BPL') model often generates better results:
\begin{equation}
f_{\rm BPL}(E) = A \left\{ \begin{array}{ll}
        (E/E_{\rm piv})^{\lambda_{l}}     &   \mbox{if $E \leq E_{\rm b}$}  \\
(E_{\rm b}/E_{\rm piv})^{\lambda_{l}} (E/E_{\rm b})^{\lambda_{h}}  &
                        \mbox{if $E > E_{\rm b}$}
\end{array}
\right. {\rm ,}
\end{equation}
where
\begin{enumerate}
\item  $A$ = amplitude in photons~s$^{-1}$~cm$^{-2}$~keV$^{-1}$,
\item  $E_{\rm piv}$ = pivot energy in keV (always held fixed),
\item  $\lambda_{l}$ = index below break,
\item  $E_{\rm b}$ = break energy in keV,
\item  $\lambda_{h}$ = index above break.
\end{enumerate}
No known physical model corresponds to this spectral function. However, a sharp 
feature in the fitted data could be a result of an absorption edge arising in the 
source. Given the likelihood of large Lorentz motions of the emitting material, 
a scenario where a sharp spectral feature could survive intact enough to be 
observable is difficult to arrange. The ambiguity of spectral deconvolution makes 
it likely that the unknown source spectrum may be less sharp than the model, while 
not allowing enough smooth curvature of the kind that characterizes the GRB function.
 
\subsubsection{The SBPL Model}

Finally, there is an intermediate model where the sharp spectral break of the 
BPL model is replaced by a smooth join of the two power law components, hence the 
name: smoothly-broken power law (`SBPL'). This model has 
the additional freedom to specify the width of the region of curvature (see also:
\cite{ryde98}):
\begin{eqnarray}
f_{\rm SBPL}(E) = A (E/E_{\rm piv})^b 10^{(\beta -\beta_{\rm piv})} &{\rm ,}  \\
\beta = m \Delta \ln{\left(\frac {e^{\alpha} + e^{-\alpha}   }  {2}\right)} & &
\beta_{\rm piv} = m \Delta \ln{\left(\frac {e^{\alpha_{\rm piv}} + e^{-\alpha_{\rm piv}}   }  
{2}\right)} \nonumber \\
\alpha = \frac {\log_{10} {(E/E_{\rm b})}} {\Delta} & &
\alpha_{\rm piv} = \frac {\log_{10} {(E_{\rm piv}/E_{\rm b})}} {\Delta} \nonumber \\
m = \frac{\lambda_2 - \lambda_1}{2} & &
b = \frac{\lambda_1 + \lambda_2}{2} \nonumber 
\end{eqnarray}
where
\begin{enumerate}
\item  $A$ = amplitude in photons~s$^{-1}$~cm$^{-2}$~keV$^{-1}$,
\item  $E_{\rm piv}$ = pivot energy in keV (always held fixed),
\item  $\lambda_{1}$ = lower power-law index,
\item  $E_{\rm b}$ = break energy in keV,
\item  $\Delta$ = break scale in decades of energy,
\item  $\lambda_{2}$ = upper power-law index.
\end{enumerate}
This model is derived as follows: we want $d(\log f) / d(\log E) = y$ (the power-law 
index as a function of energy) to be a linear function ($y = mx + b$) of the 
hyperbolic tangent  $x = \tanh \left[ \frac{\log_{10}(E/E_{\rm b})}{\Delta}\right]$, 
which interpolates smoothly between $-1$ and 1. This makes the log slope a 
horizonal line at $y=\lambda_1$ for small $E$, smoothly changing to a horizontal 
line at $y=\lambda_2$ for large $E$. For this model, the break scale parameter 
$\Delta$ is fit to its best value for the sum of all the spectra 
in the burst. For the subsequent fits to individual spectra, $\Delta$ is held 
fixed at the best-fit value, rather than being allowed to be freely determined. 
The reason for this is that $\Delta$ can not be determined better than the 
instrumental energy resolution, so the additional free parameter interferes with 
the others resulting in considerable cross-correlation that manifests itself in 
large errors. This is especially true for the individual spectral fits, where the 
statistics are worse than that of the total spectrum of the burst. Since $\Delta$ 
is held fixed throughout the series of spectral fits, the result is a model 
function with the same number of parameters of interest as the GRB function, but 
with a controllable curvature.

\subsubsection{Spectral Model Selection}

In the nominal BATSE energy band, GRBs can be successfully fit with one or more 
of these models; four parameters or fewer are sufficient. 
As can be seen from Table \ref{cat_table}, column 5, the GRB model is used most 
frequently, for 95 out of the total 156 bursts. This is followed by BPL (32 times), 
SBPL (23) and COMP (6). The actual choice of model was somewhat qualitative: if 
the GRB model did not result in a reduced $\chi^2$ distribution for all spectra 
fitted within a burst that was approximately centered on one, the BPL model was 
tried. If there was some indication from the $\chi^2_{\nu}$ distribution that the 
BPL model was inadequate, then the SBPL was used. The COMP model 
is used for those cases where the high-energy power law fit failed in more than 
$\sim 30\%$ of the individual spectra, resulting in either exponential overflow 
errors or an indeterminate value for $\beta$. 

\begin{figure*}[t!]
\centerline{\epsfig{file=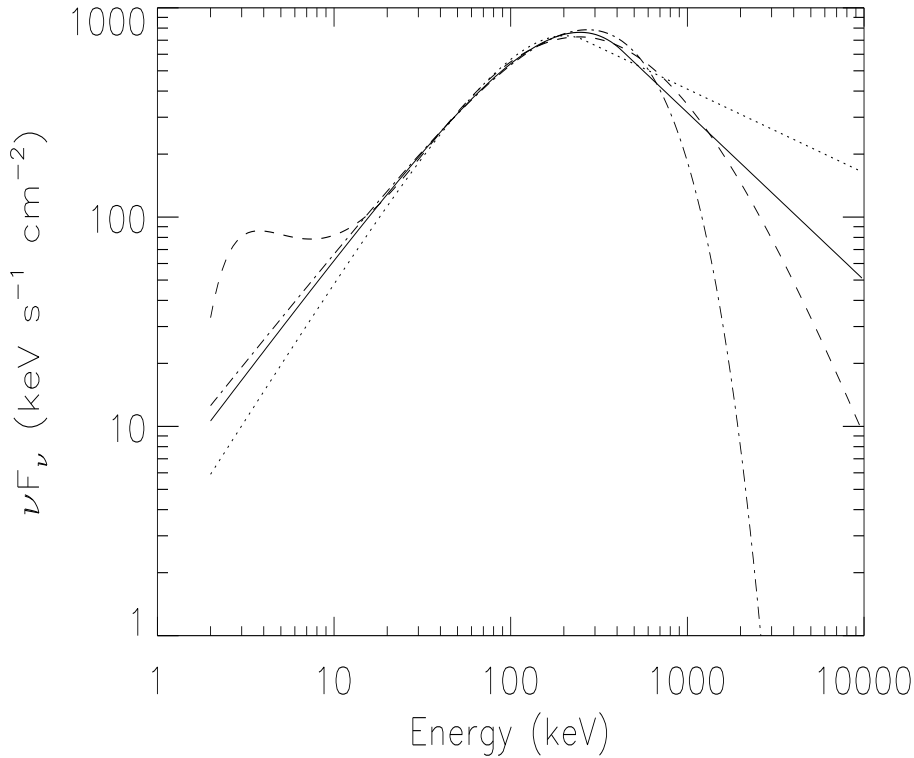,width=12cm,height=10cm}}
\caption{Several models that acceptably fit BATSE data from 
SD \#5 for the first peak of 3B 940210, 15.68--33.28 s. The models (GRB: 
{\it solid}, GRB with $\alpha$ fixed at $-2/3$: {\it dotted}, Compton Attenuation 
Model [\cite{brainerd94}]: {\it dashed} and COMP: {\it dot dashed}) are very similar 
in the energy range for which the BATSE data has good S/N, 20 to $\sim 1000$ keV, 
but diverge greatly below and above that range.
\label{extrap}}
\end{figure*}

Regardless of the model chosen, the region of applicability for the model does 
not extend beyond the energy interval fitted (see Figure \ref{extrap}). 
In fact, 
there are good reasons not to trust an extrapolation of these spectral models to 
low energies. First, at some energy around 1 keV, interstellar absorption from 
Galactic material must introduce a cut-off in the observed counts spectrum, if 
it could be measured. At slightly higher energies, but still below the LAD energy 
threshold, a considerable fraction ($15\%$) of all burst spectra have a measured 
excess in the 7--20 keV band (\cite{preece96}). This 
is corroborated by the {\it Ginga} observations of GRB spectra to energies as low
as 2 keV (\cite{stroh98}). In some of these low-energy spectra, there is also 
evidence of a low-energy spectral break distribution, complementing the well-known 
$E_{\rm peak}$ distribution (\cite{mallozzi98}). At higher energies, the picture 
is not as clear. Certainly, when it could be observed, the high energy power-law 
is seen to continue up to very high energies in the SMM, COMPTEL, EGRET, and BATSE 
SD data (\cite{matz85,hanlon95,kippen95,dingus95,briggs99b}). However, a $-2$ 
power law (consistent 
with what has been observed) can not be extended upwards in energy indefinitely, as 
it represents the emission of equal power per unit energy interval; 
there must be an energy (currently unobserved) at which the spectrum is attenuated. 
For most bursts, with high-energy power-law spectral indices $< -2$, this is not 
an issue. However, there is still the question: Where, exactly, do the 
observed non-thermal power law spectra end in energy? Eventually, broad-band 
spectroscopy may deliver an answer, but for now, it is unjustified to extend the 
fitted model spectra beyond the energies that were used in the fit.

\section{The Spectral Catalog}

We have performed spectral fitting on 156 bright bursts observed with BATSE. 
The method we use in each case starts with a preliminary spectral fit to the 
sum of all selected spectra in a burst using the GRB model. The resulting value 
of $\chi^2$ and the pattern of residuals from the model indicate whether another 
spectral model from our set should be tried, as described in the previous section. 
When the best model and parameter set for the summed spectra has been found, these 
become the starting point for a series of fits to each of the spectra in the burst. 
The average spectral parameters provide a reasonable set to fall back to in case 
one or more of the 
individual parameters is undetermined from the data. We then compare the reduced 
$\chi^2$ distribution with unity as a final check on the selected model. In some 
extreme cases, where one of the spectral parameters is undetermined in a 
large fraction of the spectra (usually the high-energy power-law component), a 
different model is tried. In the end, we have a series of spectral parameters as 
a function of time. 

The results are archived as a series of \verb+ascii+ files that are available 
electronically\footnote{On the World Wide Web: 
\verb+http://www.journals.uchicago.edu/ApJ/journal/+ (this address may change when 
the data are actually published -- there is nothing available yet)} 
having a consistent format that allows the whole series to be read by a standard 
procedure. A simple reader accompanies the archive, although the files can be 
parsed easily using any computer programming language. For the description of the 
catalog file format below, we will use as an example the file, \verb+HH_1085_4.BAT+. 
Each file's name follows a standard convention: first, the data types used in the 
analysis for the particular event are encoded at the head of the name. `HH' in the 
example represents HERB / HER, that is: 
HERB for the burst data, HER for the background; `MC' is for MER / CONT, `SS' for 
SHERB / SHER, while `CO' stands for CONT data only. The name contains the BATSE 
trigger number (see Table \ref{cat_table}) as this is a concise unique 
identifier (`1085' in the example file), whereas the burst name, `3B 911118' would 
be ambiguous for dates with more than one burst observed in the day. In Table 
\ref{cat_table}, we have added a letter that indicates the relative brightnesses 
of the several events that fall on one day (for instance, 3B 910814C [\# 676] and 
3B 910814 [\#678]). Finally, the number corresponding 
to the detector that was selected for the analysis is included (`4'), except for 
the MER files, where several detectors are summed and the detectors are 
omitted. The name always ends with the extension `\verb+.bat+'.

\subsection{Header Data}

We will reproduce portions of the example file's content below, exactly as it 
appears in the electronic archive, and describe each element in the general case. 
Each data file can be roughly divided into a standard header section and the fit 
data section. The general event information portion of the example file's header 
follows:
\begin{verbatim}
Current lookup parameters:
   Burst number:  1085
   Start day:  8578   Start sec:  68258.023
   Data types: HER HERB
   Detectors:  4

Observed selection start and stop times:
   OBS_START:  0.000000   OBS_STOP:  13.6960   51 Spectra
   Time Range:  (-7.00000, 21.0000) seconds
   Rate Range:  (2.00000, 40.0000) counts/s-keV

Selected Energy bins lower and upper thresholds (keV):
   LOWER_THRESH:  28.4235   UPPER_THRESH:  1862.51   115 Bins
   Energy Range:  (10.0000, 2000.00) keV
   Rate Range:  (0.000000, 308.096) counts/s-keV

Background selection start and stop times:
   BACK_START:  -1661.53   BACK_END:  -449.112   2 Spectra
   BACK_START:  28.8640   BACK_END:  914.856   2 Spectra
Order of the background model:  2
\end{verbatim}
The first block contains information concerning the event itself, starting with 
the BATSE trigger number: `\verb+Burst number:  1085+'. This can be used in reference 
to column 2 of Table \ref{cat_table} to find the BATSE catalog name for the event in 
column 1: `3B 911118'. The date is given in TJD: 
`\verb+Start day:  8578+', that is: Julian Day minus 2,440,000.5, and the time of 
the event trigger in seconds of day (UT): `\verb+Start sec:  68258.023+'. Following 
this, there is information about the data selection; first, the data types used, 
`\verb+Data types: HER HERB+' and the selected detector number(s): 
`\verb+Detectors:  4+'. Note that in this example `4' means detector \#4 was selected, 
{\em not} that four detectors' data were used. The data selection also gives an 
indication of the BATSE detector type, in that all data types for the SDs start 
with the letter `S' (with the one exception being `DISCSP', which is 
not used in this catalog; see Table \ref{datatypes}). Here, the HERB data 
is generated by LAD \#4, with HER for the background data. 
The next few blocks are concerned with the data selection ranges; first, the start 
and stop times with respect to the BATSE trigger time and the number of selected 
unbinned spectra: `\verb+OBS_START:  0.000000   OBS_STOP:  13.6960   51 Spectra+'. 
Next, the energy selection interval is given, with the corresponding number of data 
bins: `\verb+LOWER_THRESH:  28.4235   UPPER_THRESH:  1862.51   108 Bins+'. Finally, 
details of the background selection are presented as groups of selected spectra; 
the two statements 
`\verb+BACK_START:  -1661.53   BACK_END:  -449.112   2 Spectra+' and 
`\verb+BACK_START:  28.8640   BACK_END:  914.856   2 Spectra+' show that the background 
model consists of four spectra in two groups, one before and one after the burst 
interval (0. -- 13.69 s); the polynomial order of the background fit is 2, as seen 
in the line: `\verb+Order of the background model:  2+'. 
The `\verb+Time Range:+', `\verb+Rate Range:+' and `\verb+Energy Range:+' values define 
view windows centered on the selected time history and integrated spectrum and 
were generated as part of the data preparation step in the analysis procedure; 
as such, they have little bearing on the catalog. Taken together, all these values 
could be used to restore the data selection and views to the ones used in the 
preparation of this catalog.

The next group of information details the model used, 
and the names of the parameters.
\begin{verbatim}
          11 columns are labeled:
Broken Power Law: Amplitude (p/s-cm2-keV)
Broken Power Law: Pivot E =fix (keV)
Broken Power Law: Index < BE
Broken Power Law: Break E (kev)
Broken Power Law: Index > BE
Broken Power Law: Photon Flux (ph/s-cm^2)
Broken Power Law: Photon Fluence (ph/cm^2)
Broken Power Law: Energy Flux (erg/s-cm^2)
Broken Power Law: Energy Fluence (erg/cm^2)
Broken Power Law: Reduced Chi-squares
Reduced Chi-squares (Chi-squares in Uncertanties): 111 Degrees of Freedom
Burst Time History (Counts/s-keV)
\end{verbatim}
In this example, these are the labels of the 11 columns of the fit parameters (and 
their uncertainties, which follow) plus an additional row containing the number 
of degrees of freedom for the fit. In each label but the last, the formal model 
name: `\verb+Broken Power Law:+' comes before the parameter name: 
`\verb+Amplitude (p/s-cm2-keV)+'. The BPL model has five parameters, as discussed 
above, these make up the first five labels. Each parameter that has been fixed to 
a constant for the duration of the fit is indicated with `\verb+=fix+', as in the 
label `\verb+Broken Power Law: Pivot E =fix (keV)+'. For each model, the columns that 
follow the fit parameters contain instantaneous flux and accumulated fluence: 
`\verb+Photon Flux (ph/s-cm^2)+' and `\verb+Photon Fluence (ph/cm^2)+'. These have been 
numerically integrated over the selected energy range. In addition, corresponding 
energy fluxes and fluences have been calculated by including a factor for the energy 
in the integral: `\verb+Energy Flux (erg/s-cm^2)+' and 
`\verb+Energy Fluence (erg/cm^2)+'. The peak flux in 
photons s$^{-1}$ cm$^{-2}$ can be found in the row containing the maximum value for 
the `\verb+Photon Flux+' column. The timescale for this calculation of peak flux is, by 
necessity, the accumulation time (or time bin width) for the spectrum corresponding 
to the row containing the peak value. The total fluence is contained in the final 
row of the `\verb+Energy Fluence+' column, as this parameter records the fluence integrated 
up to including that time. Either fluence parameter can stand as a proxy to the time 
since the beginning of the burst in demonstrating relationships between fluence and 
the other fit parameters, as was done for 
$E_{\rm peak}$ in GRB peak emission (\cite{kargatis96,crider97}). Since the time 
selections and timescales may differ from the those in the official BATSE catalogs, 
these values for peak flux and fluence will almost certainly not exactly match the 
corresponding catalog values; however, they incorporate the spectral evolution 
throughout the burst, at the best available spectral resolution. The last two rows 
of data always contain the reduced $\chi^2$ values for each fit, 
`\verb+Broken Power Law: Reduced Chi-squares+', and the actual count 
history (integrated over the selected energy bins), with the label: 
`\verb+Burst Time History (Counts/s-keV)+'. In addition, we indicate the number of 
degrees of freedom for each spectral fit (111 in this case, which is constant 
throughout the burst) in a comment line that appears in the same relative position 
between the corresponding labels for reduced chi square and burst time history in 
each of the catalog data files: 
`\verb+Reduced Chi-squares (Chi-squares in Uncertanties): 111 Degrees of Freedom+'. 
The actual $\chi^2$ values can be found in the column that would 
ordinarily correspond to the uncertainty of the `\verb+Reduced Chi-squares+' column. 

Next, the selected start and stop times for the spectral accumulations are given, 
in seconds relative to the burst trigger time:
\begin{verbatim}
      50 time bins processed; start and stop times:
     0.000000     0.896000
     0.896000      1.21600
      1.21600      1.47200
      1.47200      ...
\end{verbatim}
These times bins correspond to the rows in the data arrays; the number of rows is 
given first: `\verb+50 time bins+', representing the number of {\em binned} spectra; 
thus, this number can be different from the number of unbinned spectra in the 
`\verb+OBS_START:  0.000000   OBS_STOP:  13.6960   51 Spectra+' header line. Here, two 
spectra have been combined into one to satisfy the minimum S/N requirement, reducing 
the number of time bins processed by one to 50. Usually, the start time 
of one time bin will match the stop time of the previous bin; however, in the 
case of gaps in the data, or in case quiescent periods of the burst 
were skipped, these will not be the same.

\subsection{Fit Parameters and Uncertainties}

\begin{figure*}[t!]
\centerline{\epsfig{file=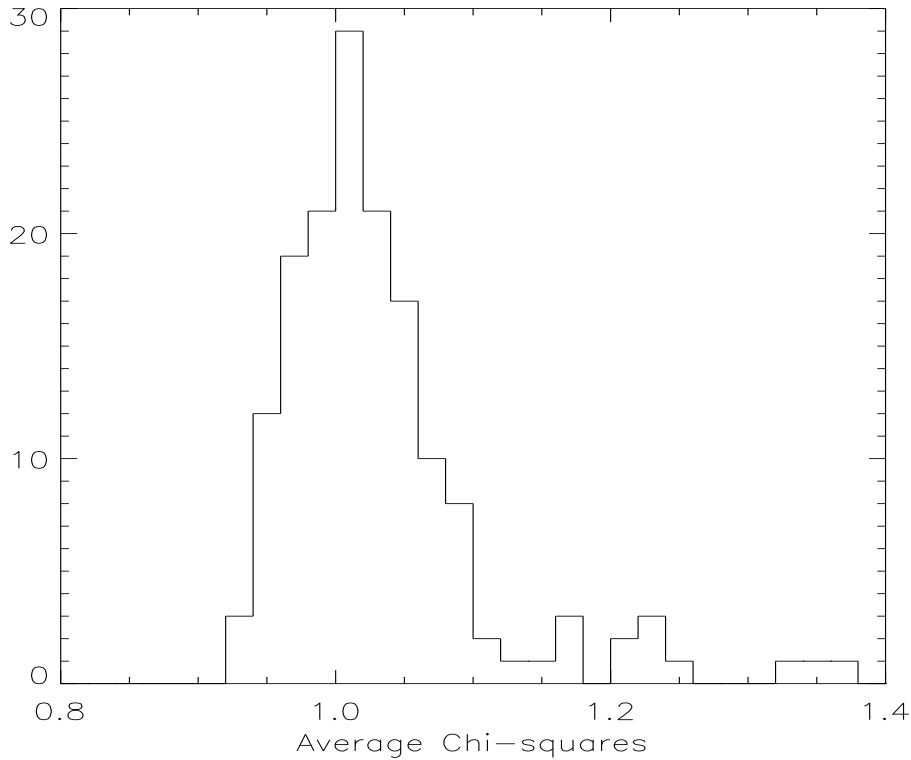,width=12cm,height=10cm}}
\caption{Total reduced $\chi^2$ distribution for the sample of 156 bursts. 
\label{chisq}}
\end{figure*}

There are two arrays of data; one each for the parameter values 
and their uncertainties. Their dimensions are listed first:
{
\begin{verbatim}
Batch Fit Parameters:       11 X      50
  0.0146409      100.000    -0.269407     326.031    -2.19617     6.27032
    5.07425  3.31890e-06  2.68582e-06    0.765253     6.70804
  0.0349295      100.000    -0.288559     284.916    -2.39224     12.6129
    8.51135  5.53011e-06  4.19282e-06     1.10751     11.3185
  0.0459018      100.000    -0.277344     280.340    -2.44527     16.5090
    11.9836  7.19965e-06  5.70708e-06    0.940453     14.0348
  0.0592305      100.000    -0.285810     247.583    -2.43234     18.5416
    14.8440  6.65124e-06  6.73318e-06     1.27628     16.1302
  0.0757569      100.000    -0.305579     201.955    -2.13975     22.2641
   ...
\end{verbatim}}
In this case, there are eleven columns that correspond to the eleven column labels 
discussed above. There are 50 rows of data, corresponding to the 50 time bins of 
accumulated spectra. In this representation, each row has been split into two, 
to accommodate the eleven columns of data. A single column of this table represents 
the time history of the corresponding fit parameter (or other value, as indicated 
above) over the time history of the burst. For comparison, the count rate history, 
summed over the selected energy bins, is found in the last column.
Notice that the column corresponding to the parameter that was fixed 
(`\verb+Broken Power Law: Pivot E =fix (keV)+') has all identical rows, as appropriate 
for a parameter held constant throughout. Also, the seventh and ninth columns
are monotonically increasing, as these correspond to the accumulated fluences. 
When the reduced $\chi^2$ values (in the tenth column) are displayed as a histogram 
they should be distributed fairly uniformly around 1.0 for an acceptable series of 
fits. Any deviation of the centroid of the distribution from 1.0 by more 
than $3 \sigma$ (as defined by the HWHM of the distribution) indicates possible 
systematic effects. Figure \ref{chisq} shows the histogram of centroids for the 
individual reduced $\chi^2$ distributions for the entire catalog. 
Some bursts are so intense in individual spectra that they belong to the regime 
where the systematics inherent in our model of the detector electronics and 
response dominate over the counts statistics, and so these have unusually 
high $\chi^2$ values.

The final block of data lists the uncertainties for each of the fitted parameters 
(except that $\chi^2$ is given in the place of an uncertainty for the reduced 
$\chi^2$, as discussed above). The uncertainties are given as the $1 \sigma$ errors 
derived from the covariance matrix of the fit. The given error estimate assumes the 
derived uncertainty is symmetrical about the best fit value of the parameter, which is 
good to first order, where each parameter is considered singly. To know the error 
distribution in more detail requires creating a map of how $\chi^2$ changes near the 
fitted parameter value. As the result depends upon the number of total parameters that 
are of interest for the particular question being asked, such analyses are 
not appropriate for a general-use catalog. Note that the fitting process may 
not have been able to determine the value of some parameters from the data. Such 
cases are flagged as indeterminate with a zero in the uncertainty. The fitting 
routine automatically fixes the value of an indeterminate parameter to the 
best-fit value determined for the integrated spectrum, while 
it attempts to optimize the fit for the remaining parameters for that particular 
spectrum. The parameter is allowed to vary again as usual when the next spectrum 
in the time series is fit.

In some cases, there may be correlations between parameters that are not evident 
from their individual distributions over each event. 
These may be recovered by observing a correlation in the scatter plot of the two 
parameters. Some of these correlations may arise from physical processes associated 
with GRB emission while some may be related to the statistics of parameter 
estimation when the two parameters are considered jointly for the chosen model. 
In the latter case, a more detailed analysis of the parameter uncertainties will have 
to be made. This was done, for example, for the low-energy power-law spectral 
index $\alpha$, obtained by either the GRB or the BPL model, to make a direct 
comparison with a specific prediction associated with a particular model 
(\cite{preece98b}). Simulations of the error distribution for different values 
of $\alpha$ indicated that larger (less negative) values had systematically 
larger errors associated with them, as was indeed observed.

\section{Results}

\subsection{Distributions of Spectral Parameters}

\begin{figure*}[t!]
\centerline{\epsfig{file=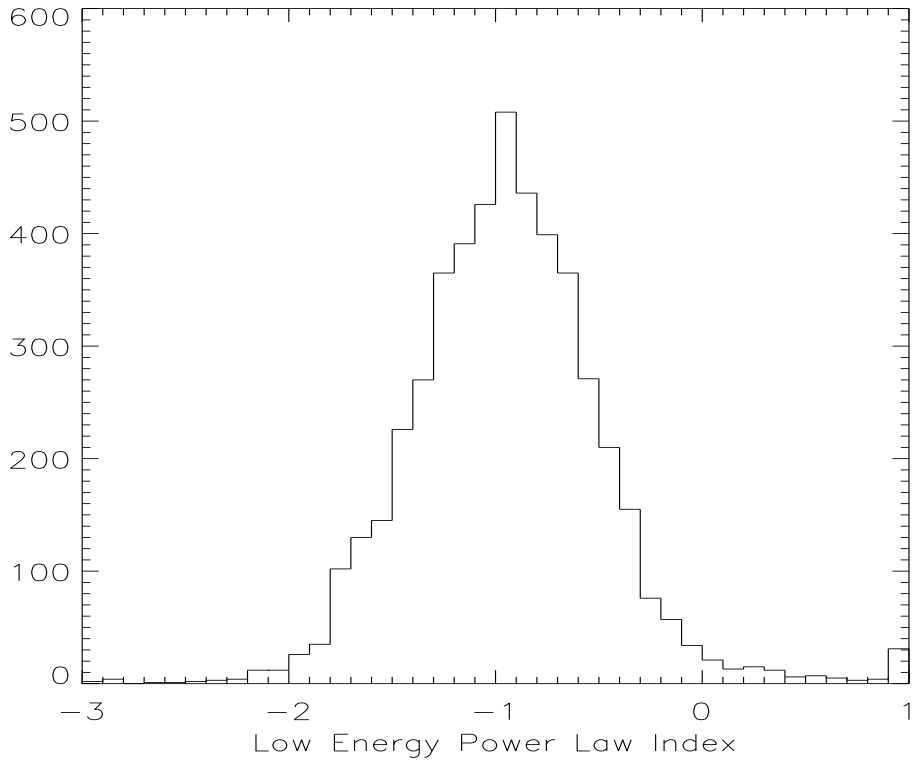,width=12cm,height=10cm}}
\caption{Low-energy power law index distribution for the entire sample. 
\label{alpha_dist}}
\end{figure*}

\begin{figure*}[t!]
\centerline{\epsfig{file=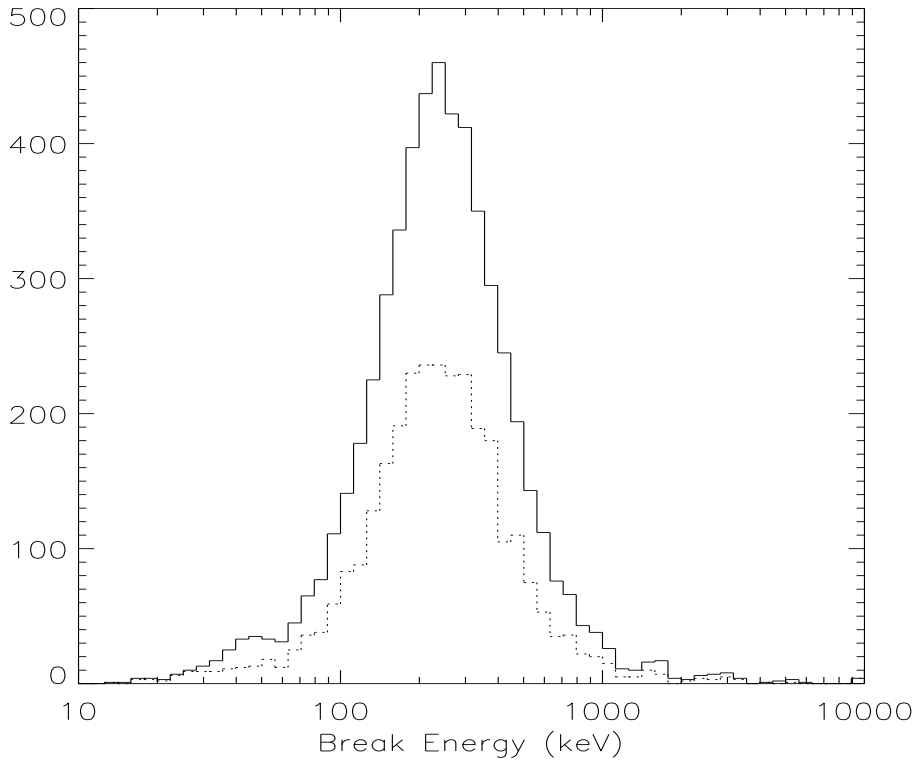,width=12cm,height=10cm}}
\caption{Break energy distribution for the entire sample ({\it 
solid line}). The subset of bursts {\em not} fit with the GRB model is also shown 
({\it dotted line}).
\label{ebreak_dist}}
\end{figure*}

To get a general idea of the contents of the catalog, we present here the overall 
distributions of the three spectral parameters that have equivalents in each of the 
spectral models we have used for fitting (see Figures \ref{alpha_dist}, 
\ref{ebreak_dist} \& \ref{beta_dist}). Each distribution is made up of the fitted 
parameters from 5500 different spectra, except for the high-energy power-law 
index distribution, which has 5253, since $\beta$ sometimes was omitted from the fits. 
This occurred either because the parameter was absent from the model fitted (as in 
the COMP model) or because it was ill-determined and so it is fixed in value as 
discussed above. This will happen for spectra 
where there are not enough high-energy counts to constrain the value. 

The low-energy power-law index $\alpha$ distribution is seen not to overlap 
significantly with that of the high-energy ($\beta$; see Figures \ref{alpha_dist} 
\& \ref{beta_dist}). Even though the distribution of $\alpha$ falls sharply on 
the high end toward $+1$, the edge is not as sharp as the expected statistical 
error, assuming that all bursts have that same value of $-1$ (or even $-2/3$, 
as expected from single-particle synchrotron emission 
-- \cite{preece98b}). That being said, it is interesting that the most likely value 
for $\alpha$ is $-1$, not $-2/3$, so this number should arise naturally out of the 
physics of the emission mechanism. So far, an explanation that requires this value 
has not surfaced. However, the spread in values is also significant, 
especially those that are greater than $-2/3$, indicating bursts where synchrotron 
emission alone is not an option. 

Spectral evolution within many of the bursts analyzed here has two consequences: 
First, the dataset consists of series of parameters, some of which appear to be 
correlated from one spectrum to the next, so the total distributions presented 
here consist of points that may not seem to be statistically independent. However, 
spectral evolution within a burst is defined by the 
observation that some spectral shape parameter is changing over the history of 
the burst, implying there is an intrinsic spread to the distribution related to 
the evolution of the parameter. That being said, it is still remarkable that the 
spectral break parameter, which has been the single most important signifier for 
spectral evolution (\cite{ford95}), has a distribution (Figure \ref{ebreak_dist})  
with a half-maximum full width less than a decade in energy. 
The significance of 
this is all the more emphasized by the fact that, out of all three spectral shape 
parameters, the break energy is the only one that is sensitive to relativistic 
motion between the source and the detector. On the one hand, cosmological 
red-shifting of GRB sources has been inferred from the shifts between break energy 
distributions for different peak flux bands (\cite{mallozzi95}). In the present 
work, most of the bursts selected could be considered to belong to a single 
intensity group, so this effect should not be present (see Figure \ref{lognlogp}). 
However, given that bursts are now known to have 
considerable spreads in their intrinsic luminosities, it is highly likely that 
the bursts making up the sample set come from a large variety of distances, with 
an associated spread in red-shifts. This is clearly not evident in the observed 
distribution; the spread from $z$ could easily be a factor of a few, but that would 
imply a correspondingly narrower intrinsic distribution. On the other hand, 
the presence of very high energy photons in many bursts implies very high Lorentz 
boosting of the emitting regions (\cite{matz85,hurley94}) in the blast wave scenario 
(\cite{goodman86,meszaros93,reesmesz94}). It has been noted by \cite{brainerd94} 
that, despite a sensitive dependence upon the Lorentz boost factor, the break energy 
distribution does not seem to fill the pass band of the detectors, rather it is 
confined mainly to within a single decade of energy, as seen clearly in Figure 
\ref{ebreak_dist}. Detailed modeling of the threshold detection efficiency for 
bursts with high and low break energies (\cite{brain99}), as well as an extensive 
search yielding negative results for a class of `high break energy' bursts 
undetectable by BATSE in the SMM data (\cite{harris98}), together indicate the 
importance of the narrowness of the distribution as a clue to the source emission 
mechanism. 

\subsection{General Discussion}

There appears to be a second population of low-energy spectral break bursts, as 
evidenced by the second, smaller peak in Figure \ref{ebreak_dist} centered at 
$\sim 45$ keV. Although the energy is not coincident with the lowest energy available 
to be fitted, it may be that the grouping of low-energy breaks near the lowest 
available energy represents a `pinning' of the parameter at a low value, whereas 
the spectrum actually does not significantly change slope below that value. 
The GRB model has a large intrinsic curvature below the break energy that is 
related to how the spectrum smoothly joins to a low-energy power-law component. 
Depending on the difference $\alpha - \beta$, which is typically close to 1, this 
fixed-curvature portion of the function can span several tens of keV above the lowest 
available energy. This can be tested by examining the distribution that results 
when the fitted values derived from the GRB model have been removed from the data 
set. In this case, most of the second peak vanishes, as seen in the dotted 
histogram in Figure \ref{ebreak_dist}. There is still the open question of 
the reality of a second, low-energy, continuum spectral break in some GRBs, since 
there is tantalizing evidence for this from the {\it Ginga} data set (\cite{stroh98}). 
Most of the bursts in the {\it Ginga} sample have fitted $E_{\rm break}$ values that 
cluster around the peak of the BATSE distribution (Figure \ref{ebreak_dist}), 
however, 6 out of the 22 total have values below 10 keV. BATSE SD discriminator 
data has the potential to resolve some of this issue by extending spectral fits 
to BATSE data to lower energies (although not below 8 keV), indicating that four 
out of 86 bursts fall into the category of low-energy spectral breaks 
(\cite{preece96}).

\begin{figure*}[t!]
\centerline{\epsfig{file=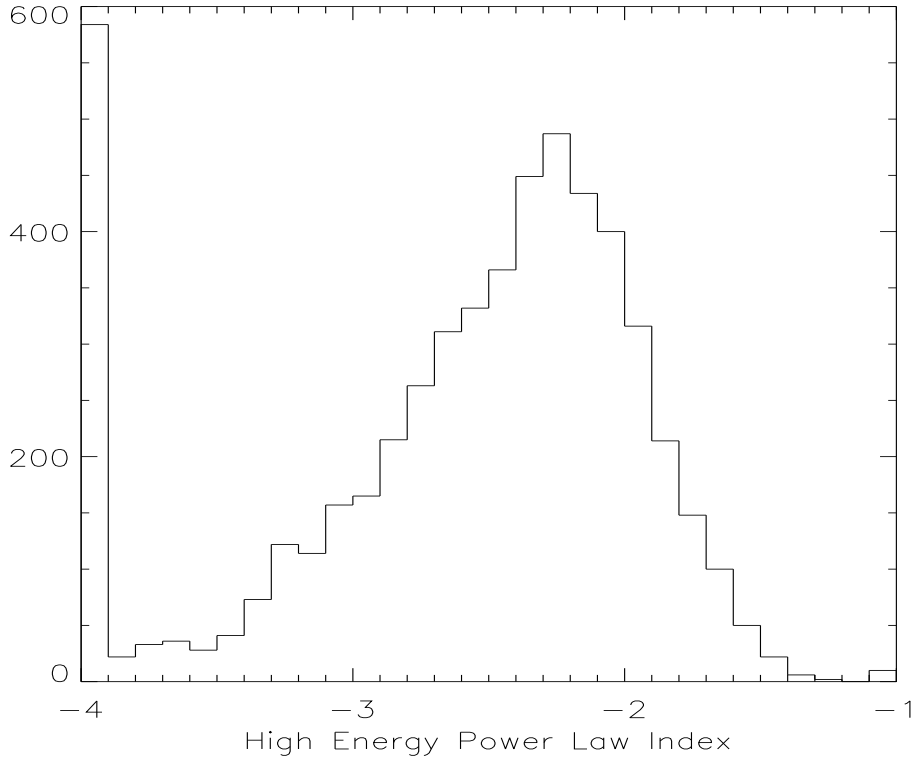,width=12cm,height=10cm}}
\caption{High-energy power law index distribution for the entire 
sample. All spectra that either do not have $\beta$ values or that have values 
$<-4$ have been included in the lowest bin.
\label{beta_dist}}
\end{figure*}

The general results from previous work on GRB spectroscopy are confirmed and 
expanded in the current database. A spectral form with two power law segments 
connected by some curvature seems to be the most prevalent acceptable model 
(\cite{band93}). However, the COMP model, with no high-energy power law portion, 
was found to be an acceptable fit to all spectra in a single burst for six bursts 
out of the total sample. This may reflect the broad classification of bursts into 
high-energy (HE) and no-high energy (NHE) (\cite{pendleton97}). There, as here, 
the phenomenon is characterized by a somewhat arbitrary division of a smooth 
distribution of high-energy behaviors into two groups. Here, we used the COMP 
model for those cases where the high-energy power-law index was poorly 
constrained in an initial global fit to the sum of all selected spectra within a 
burst. We took a value of the high-energy power-law index of $< -4$ 
to indicate a poor constraint, since in practice many of the spectra fitted 
individually in such a case resulted in a poorly-determined value, or worse, a 
math underflow error in the function evaluation. Despite this index cut for the 
global spectrum in each burst, many individual spectral fits result in very 
large negative values for the index, which have been added to the lowest bin 
in Figure \ref{beta_dist}. 
Altogether there are 308 spectra with high-energy power-law indices below $-4$, in 
addition to the 247 spectra from the six bursts fit with the COMP model, or 10\% of 
the spectra fitted that effectively do not have a high-energy power-law component. 
This is also in accordance with the results that NHE portions exist within HE 
bursts (\cite{pendleton97}). There are differences, in that here, we do not take 
into account the effect of the break energy on whether a burst has high-energy 
emission, since, if the break energy is sufficiently high, the high-energy portion 
of the burst may not be well determined. 

\subsection{Concluding Remarks}

It is our hope that this data set will continue to be useful for correlative 
analyses and testing of theoretical predictions. It has already 
generated some results of this nature, based on preliminary versions of the catalog. 
The peak of the high-energy power-law distribution is the same as reported before, 
$-2.25$ (\cite{preece98a}), although now with much better statistics. 
This is expected, as the data set used previously is a subset of the current one. 
In the previous work, the distribution for the average of the high-energy 
power-law index over each burst was presented, while here, we have shown the entire 
time-resolved distribution. Crider et al. (1997) 
have looked into bursts with unusually high values for the low-energy power-law 
index, as a possible indicator for saturated synchrotron self-Compton in isolated 
peaks of emission. The overall distribution of low-energy power-law indices has 
lead to a challenge to the validity of the synchrotron shock model 
(\cite{katz94,tavani95}). Indeed, the distribution of $\alpha$ is centered on $-1$, 
which should be accounted for by any valid model. It is also too wide to accommodate 
the strict limit of $-2/3$ that is imposed by the synchrotron shock model. The 
narrowness of the energy break distribution is also important. Many other studies 
can be constructed that correlate parameters within individual bursts, for example, 
or study average behavior over the entire population. 

Following the present work, there will be two additional spectroscopy catalogs, one 
with fits to the peak flux and integrated fluence spectra using medium energy 
resolution BATSE data (\cite{bob_cat}), and the other presenting direct deconvolved 
spectra using four-channel discriminator data for throughout each burst 
(\cite{geoff_cat}). 
These catalogs will be concerned with the spectroscopy for larger burst samples 
than we have been able to analyze here, given our requirements. The medium energy 
resolution data can provide spectral parameters similar to those presented in the 
current work for the peak flux and total fluence spectra for most bursts in the 
BATSE catalog. For very weak 
bursts at the detector threshold, data with very coarse energy resolution may be 
employed to determine spectral hardness parameter evolution for every event. 

\acknowledgments

Thanks to the BATSE operations team for some timely peak flux and fluence calculations, 
especially Surasak Phengchamnan, Burl Peterson and Maitrayee Sahi.



\end{document}